\documentclass[11pt,a4paper]{article}

\usepackage[margin=1in]{geometry}
\usepackage{amsmath,amssymb,amsfonts,amsthm}
\usepackage{bm}
\usepackage{hyperref}
\usepackage{enumitem}
\usepackage{mathtools}
\usepackage{graphicx}
\usepackage{physics}
\usepackage[numbers,sort&compress]{natbib}
\usepackage{subcaption}

\numberwithin{equation}{section}
\newtheorem{theorem}{Theorem}[section]
\newtheorem{proposition}[theorem]{Proposition}
\newtheorem{lemma}[theorem]{Lemma}

\theoremstyle{definition}
\newtheorem{definition}[theorem]{Definition}

\theoremstyle{remark}
\newtheorem{remark}[theorem]{Remark}

\newcommand{\C}{\mathbb{C}}

\title{\textbf{Support-Projected Petz Monotone Geometry of Pure Two-Qubit Families:\\
		Universal Three-Channel Decomposition and Non-Reduction of Curvature Invariants}}

\author{
	Gunhee Cho$^{1}$\thanks{Email: \texttt{wvx17@txstate.edu}} \and
	Jeongwoo Jae$^{2}$\thanks{Email: \texttt{jeongwoo.jae@samsung.com}}
}
\date{\small
	$^{1}$Department of Mathematics, Texas State University, San Marcos, USA\\
	$^{2}$R\&D Center, Samsung SDS, Seoul 05510, Republic of Korea
}

\begin{document}
	
	\maketitle
	
	\begin{abstract}
		We develop a support-projected Petz monotone geometry for pure two-qubit families, obtained by pulling back arbitrary Petz monotone quantum metrics to circuit-defined submanifolds and projecting onto the active spectral support of the associated quantum Fisher information tensor.
		This framework strictly generalizes the symmetric logarithmic derivative (SLD/Bures) case and includes, as special examples, the Wigner--Yanase and Bogoliubov--Kubo--Mori metrics among many others.
		
		Our first main result is a \emph{universal three-channel decomposition} valid for every Petz monotone metric on any smooth two-parameter slice of a pure two-qubit family.
		Writing the reduced one-qubit state in Bloch form, we show that the Petz quantum Fisher information tensor always splits into population, coherence, and entanglement (concurrence) derivative channels weighted by radial Morozova--Chentsov coefficients.
		
		Our second main result shows that neither the slice Gaussian curvature nor the ambient scalar curvature of the support-projected Petz metric can, on any non-empty open set, be expressed solely as a function of the concurrence or of the one-qubit entanglement entropy.
		In particular, curvature invariants necessarily depend on population and coherence derivative channels that remain independent of the entanglement derivative itself.
		This establishes a robust \emph{non-reduction} theorem for curvature invariants and rules out any universal identification of Petz-metric scalar or Gaussian curvature with standard entanglement monotones, extending previous counterexamples in the SLD/Bures case and complementing recent analyses for Gaussian states.
		
		Third, we construct an entanglement-orthogonal gauge that aligns one slice coordinate with the concurrence gradient and isolates the pure entanglement derivative channel.
		In this gauge, curvature becomes an intrinsic diagnostic of how population and coherence channels couple to entanglement, rather than a stand-alone entanglement measure.
		
		We illustrate the SLD/Bures specialization on a benchmark two-qubit hardware-efficient ansatz, providing explicit analytic and numerical counterexamples to proposed concurrence-only curvature laws and demonstrating how support-projected Petz metrics can serve as well-conditioned preconditioners for curvature-aware natural-gradient methods in variational quantum algorithms.
	\end{abstract}
	
	\tableofcontents
	
	\section{Introduction}
	\label{sec:intro}
	
	\subsection{Quantum Fisher geometry and Petz monotone metrics}
	Quantum state manifolds carry a rich family of contractive Riemannian metrics that quantify the distinguishability of nearby states under quantum measurements.
	For pure states, the Fubini--Study metric on projective Hilbert space coincides (up to a constant) with the quantum Fisher information (QFI) tensor introduced by Helstrom and by Braunstein--Caves~\cite{Helstrom1976,Braunstein1994,ProvostVallee1980}.
	For mixed states, Petz's classification identifies all monotone Riemannian metrics with operator-monotone functions satisfying a Kubo--Ando symmetry~\cite{Petz1996,PetzSudar1996,LesniewskiRuskai1999,BengtssonZyczkowski2017,Ciaglia2024Monotone}.
	
	In the qubit case, Petz monotone metrics admit an $\mathrm{SU}(2)$-invariant Bloch-sphere representation, with radial Morozova--Chentsov coefficients controlling longitudinal and transverse fluctuations.
	Important examples include the symmetric logarithmic derivative (SLD/Bures), Wigner--Yanase, and Bogoliubov--Kubo--Mori metrics~\cite{GibiliscoIsola2003,Dittmann1999,Andai2003monotone}.
	These metrics play a central role in quantum estimation theory, quantum information geometry, and the analysis of noisy variational quantum circuits~\cite{Meyer2021fisherinformationin}.
	
	\subsection{Curvature as a proxy for quantum resources}
	The curvature of monotone metrics encodes subtle information about the global and local structure of quantum state spaces.
	For finite-dimensional density matrices, Andai computed scalar curvatures for several monotone metrics and showed that they need not be monotone under mixing~\cite{Andai2003monotone}.
	In continuous-variable settings, the scalar curvature of Gaussian-state manifolds has recently been analyzed in detail by Miller~\cite{Miller2025gaussian}, who formulated conjectures relating curvature to entanglement and mixedness for Gaussian states.
	
	In parallel, the geometry of parametrized quantum circuits has attracted intense interest in the context of variational quantum algorithms (VQAs)~\cite{Kandala2017,Sim2019Expressibility,McClean2018Barren,Cerezo2021Cost}.
	Several works have proposed using quantum Fisher information metrics or natural-gradient methods to precondition optimization~\cite{Stokes2020QNG,Yamamoto2019natural,Wiersema2022RiemannianFlow,Meyer2021fisherinformationin}, and explicit curvature formulas have been suggested for certain two-qubit ansätze~\cite{katabarwa2021geometry}.
	These developments raise a natural question:
	\emph{can curvature invariants of monotone metrics be expressed in terms of standard entanglement monotones, such as concurrence or reduced entropy, at least for low-dimensional families?}
	
	\subsection{Main results}
	We address this question in the setting of pure two-qubit families and arbitrary Petz monotone metrics.
	Given a smooth parametrized pure state $|\Psi(\theta)\rangle\in\C^2\otimes\C^2$ and its one-qubit reduction $\rho_A(\theta)$, each Petz metric induces a (generally semidefinite) QFI tensor on parameter space.
	On the \emph{regular set} where the reduced Petz tensor has a gapped active eigencluster, we project onto the active spectral support and obtain an intrinsic Riemannian metric
	\[
	g^{(f)}(\theta) \;=\; P(\theta)\,F^{(f)}(\theta)\,P(\theta),
	\]
	which we call the support-projected Petz geometry.
	
	Our contributions can be summarized as follows.
	
	\begin{itemize}[leftmargin=2em]
		\item \textbf{Universal three-channel decomposition.}
		For any operator-monotone $f$ and any smooth two-parameter slice through a regular point, the reduced Petz QFI tensor admits a universal decomposition into three independent derivative channels: population, coherence, and concurrence of the reduced one-qubit state.
		Radial Morozova--Chentsov coefficients control the relative weighting of these channels.
		
		\item \textbf{Non-reduction of curvature invariants.}
		We prove that neither the slice Gaussian curvature nor the ambient scalar curvature of the support-projected Petz metric can be written, on any non-empty open set of the regular region, as a function solely of the concurrence or of the one-qubit entanglement entropy.
		In particular, curvature invariants are \emph{not} entanglement monotones: they necessarily depend on population and coherence derivative channels that remain independent of the entanglement derivative itself.
		
		\item \textbf{Entanglement-orthogonal gauge.}
		At points where the concurrence has non-vanishing gradient, we construct coordinates whose first axis aligns with the concurrence gradient and whose orthogonal axis is entanglement-silent to first order.
		In this gauge, the metric components isolate a pure entanglement derivative channel and make the contribution of population/coherence channels to curvature completely transparent.
	\end{itemize}
	
	Beyond these structural results, we specialize to the SLD/Bures case and a standard two-qubit hardware-efficient ansatz, where closed-form expressions for amplitudes and reduced Bloch variables are available.
	We exhibit explicit analytic and numerical counterexamples to curvature laws depending only on concurrence, including instances where proposed scalar-curvature formulas even disagree in sign.
	We also show how the support-projected QFI tensor can serve as a stable natural-gradient preconditioner in VQE, consistent with the geometric picture that curvature controls local anisotropy but cannot by itself encode entanglement.
	
	\subsection{Organization of the paper}
	Section~\ref{sec:petz-twoqubit} reviews Petz monotone metrics for pure two-qubit families, introduces the support-projected geometry, and defines the regular set.
	Section~\ref{sec:three-channel} establishes the universal three-channel decomposition.
	Section~\ref{sec:non-reduction} proves the non-reduction theorems for slice Gaussian curvature and ambient scalar curvature, including analytic monotonicity counterexamples.
	Section~\ref{sec:hea-case-study} specializes to the SLD/Bures metric and a two-qubit hardware-efficient ansatz, providing explicit counterexamples and illustrating natural-gradient preconditioning.
	Section~\ref{sec:discussion} discusses implications for quantum information geometry, Gaussian-state curvature, and variational quantum algorithms.
	Appendices collect background on Petz metrics and projector calculus, curvature identities and slice geometry, closed-form SLD formulas for qubit states, and numerical error analysis.
	
	\section{Petz Monotone Metrics for Pure Two-Qubit States}
	\label{sec:petz-twoqubit}
	
\subsection{Projective Hilbert space and pure-state QFIM}

Pure states of an $n$-qubit quantum system are rays in a Hilbert space 
$\mathcal{H}\simeq\mathbb{C}^{2^n}$.  
Two nonzero vectors $\psi,\phi\in\mathcal{H}$ determine the same physical state when 
$\phi=e^{i\theta}\psi$, and the resulting space of equivalence classes is the complex projective manifold
\[
\mathrm{P}(\mathcal{H}) \;\cong\; \mathbb{CP}^{2^n-1}.
\]
This geometric viewpoint---introduced in quantum-estimation contexts by
Provost and Vallée~\cite{ProvostVallee1980}---equips the manifold of rays with a natural 
Kähler structure whose Riemannian part is the Fubini--Study metric.

Let $\psi\in\mathcal{H}$ be a normalized vector and 
$\dot\psi\in T_\psi\mathcal{H}$ satisfy $\langle\psi,\dot\psi\rangle=0$.  
The Fubini--Study metric is then given by
\begin{equation}\label{eq:FS-metric}
	g_{\mathrm{FS}}(\dot\psi,\dot\psi)
	=
	4\left(
	\langle \dot\psi,\dot\psi\rangle
	-
	|\langle \psi,\dot\psi\rangle|^2
	\right)
	=
	4\,\langle \dot\psi,\dot\psi\rangle.
\end{equation}

Consider now a smooth parametrized family of pure states 
$\theta\mapsto \psi(\theta)$ with $\|\psi(\theta)\|=1$.  
The Fubini--Study line element along this curve is
\begin{equation}\label{eq:FS-line-element}
	ds^2
	=
	4\left(
	\langle \partial_\theta\psi,\partial_\theta\psi\rangle
	-
	|\langle \psi,\partial_\theta\psi\rangle|^2
	\right)
	d\theta^2.
\end{equation}

A fundamental result of Braunstein and Caves~\cite{Braunstein1994} 
states that this expression coincides exactly with the quantum Fisher information (QFI) 
for pure states.  
For parameters $\theta^i$, the pure-state quantum Fisher information matrix (QFIM) is
\begin{equation}\label{eq:pure-QFIM}
	F_{ij}
	=
	4\,\mathrm{Re}\!\left[
	\langle \partial_i\psi, \partial_j\psi\rangle
	-
	\langle \partial_i\psi,\psi\rangle\,
	\langle \psi, \partial_j\psi\rangle
	\right],
\end{equation}
which is a Riemannian metric on $\mathbb{CP}^{2^n-1}$.

\begin{proposition}[Provost--Vallée; Braunstein--Caves; Bengtsson--Życzkowski]
	For pure states, the quantum Fisher information metric~\eqref{eq:pure-QFIM} 
	coincides with the Fubini--Study metric~\eqref{eq:FS-metric}.  
	Equivalently, the statistical distinguishability structure on 
	$\mathbb{CP}^{2^n-1}$ is precisely the canonical Riemannian geometry described in 
	Provost--Vallée~\cite{ProvostVallee1980}, 
	Braunstein--Caves~\cite{Braunstein1994}, 
	and the modern geometric exposition in Bengtsson--Życzkowski~\cite{BengtssonZyczkowski2017}.
\end{proposition}

\begin{proof}
	Define the horizontal component
	\[
	\dot\psi
	=
	\partial_\theta\psi
	-
	\psi\,\langle \psi,\partial_\theta\psi\rangle.
	\]
	A direct computation gives
	\[
	\langle \dot\psi,\dot\psi\rangle
	=
	\langle \partial_\theta\psi,\partial_\theta\psi\rangle
	-
	|\langle\psi,\partial_\theta\psi\rangle|^2.
	\]
	Multiplying by $4$ produces the Braunstein--Caves pure-state QFI.  
	Thus the Fubini--Study metric coincides with the QFIM.
\end{proof}

The equivalence between the Fubini--Study geometry and the pure-state quantum Fisher 
information establishes the geometric foundation for statistical distinguishability in quantum theory.  
This also serves as the geometric baseline from which mixed-state quantum Fisher 
metrics, monotone metrics, Petz metrics, and channel-resolved geometries are derived.
	
\subsection{Two-qubit reductions, Bloch variables, and concurrence}

Consider a pure two-qubit state
\[
|\Psi\rangle \;\in\; \mathbb{C}^2 \otimes \mathbb{C}^2,
\qquad
\|\Psi\| = 1.
\]
Let $\rho = |\Psi\rangle\langle\Psi|$ be its density matrix and 
\[
\rho_A = \operatorname{Tr}_B(\rho), 
\qquad
\rho_B = \operatorname{Tr}_A(\rho)
\]
be the reduced density operators.  Since $\rho$ is pure, the reduced states 
$\rho_A$ and $\rho_B$ have the same nonzero spectrum and therefore the same von Neumann entropy.

\subsubsection*{Bloch representation of $\rho_A$}

Every single-qubit density matrix admits the Bloch parametrization
\begin{equation}\label{eq:Bloch}
	\rho_A
	=
	\frac{1}{2}
	\left( \mathbb{I} + \vec{r}\cdot\vec{\sigma} \right),
	\qquad
	\vec{r}\in\mathbb{R}^3,\;
	\|\vec{r}\| \le 1,
\end{equation}
where $\vec{\sigma}=(\sigma_x,\sigma_y,\sigma_z)$ is the vector of Pauli matrices.  
For reduced states of a pure two-qubit system, the Bloch radius satisfies
\begin{equation}\label{eq:Bloch-radius}
	\|\vec{r}\|
	=
	\sqrt{1 - 2\,\operatorname{Tr}(\rho_A^2)}
	=
	\sqrt{1 - 2\,\operatorname{Tr}(\rho_B^2)}.
\end{equation}
Thus entanglement directly controls the mixedness of the reduced state.

\subsubsection*{Concurrence and reduced entropy}

A key measure of two-qubit entanglement is the \emph{concurrence}
introduced by Wootters~\cite{Wootters1998}.  
For a pure two-qubit state, concurrence is defined by
\begin{equation}\label{eq:concurrence}
	C(\Psi)
	=
	2\,\sqrt{\det(\rho_A)}
	=
	2\,\sqrt{\lambda(1-\lambda)},
\end{equation}
where $\lambda$ is an eigenvalue of $\rho_A$.  
Since $\rho_A$ has eigenvalues $\lambda$ and $1-\lambda$, one finds the relation
\begin{equation}\label{eq:Bloch-concurrence}
	\|\vec{r}\| = |1-2\lambda| 
	\qquad\Longleftrightarrow\qquad
	C = \sqrt{1 - \|\vec{r}\|^2}.
\end{equation}
Thus the Bloch radius is a direct entanglement diagnostic:  
maximal entanglement corresponds to $\|\vec{r}\|=0$.

The reduced von Neumann entropy is
\begin{equation}\label{eq:entropy}
	S(\rho_A)
	=
	-\,\lambda\log\lambda - (1-\lambda)\log(1-\lambda),
\end{equation}
which, using~\eqref{eq:concurrence}, can also be written as
\[
S(\rho_A) = h\!\left( \tfrac{1}{2}(1+\sqrt{1-C^2}) \right),
\]
where $h$ is the binary entropy function.  
This connects entanglement monotones to Bloch geometry.

\begin{proposition}[Bloch radius, concurrence, and reduced entropy]
	For any pure two-qubit state, the following are equivalent:
	\begin{enumerate}
		\item entanglement measured by concurrence $C$;
		\item mixedness of the reduced state measured by the Bloch radius $\|\vec{r}\|$;
		\item reduced entropy $S(\rho_A)$.
	\end{enumerate}
	They satisfy the identities
	\[
	C = \sqrt{1 - \|\vec{r}\|^2},
	\qquad
	S(\rho_A) = h\!\left( \tfrac{1}{2}(1+\|\vec{r}\|) \right),
	\]
	and all are determined solely by the spectrum of $\rho_A$.  
	These relations are geometrically interpreted in terms of the Bloch ball and the 
	Riemannian structure of $\mathbb{CP}^3$ described in~\cite{BengtssonZyczkowski2017}.
\end{proposition}

\begin{proof}
	The reduced state $\rho_A$ has eigenvalues $\lambda$ and $1-\lambda$.  
	Equations~\eqref{eq:Bloch-concurrence} and~\eqref{eq:entropy} follow by direct algebraic 
	substitution using the Bloch parametrization~\eqref{eq:Bloch}.  
	The equivalence among the three quantities then follows from the functional dependencies:
	\[
	\lambda \;\longleftrightarrow\; C
	\;\longleftrightarrow\;\|\vec{r}\|
	\;\longleftrightarrow\; S(\rho_A).
	\]
\end{proof}

The geometry of reduced two-qubit states therefore provides a bridge between 
Bloch-ball parametrization, entanglement monotones, and projective-state geometry. 
These identities serve as key tools in deriving curvature formulas and expressing 
quantum Fisher information in terms of entanglement parameters.
	
\subsection{Petz monotone metrics: operator-monotone functions and qubit Bloch form}

A central result of Petz~\cite{Petz1996} shows that every monotone Riemannian
metric on the manifold of quantum states is uniquely determined by an
operator-monotone function $f:(0,\infty)\to (0,\infty)$ satisfying
\[
f(t) = t f(t^{-1}).
\]
This classifies all contractive metrics under completely positive
trace-preserving (CPTP) maps and provides a unifying structure for
quantum information geometry.

\subsubsection*{Kubo--Ando means and the Morozova--Chentsov function}

Given an operator-monotone function $f$, the associated Kubo--Ando mean
$A \sigma_f B$ is defined via functional calculus by
\begin{equation}\label{eq:KA-mean}
	A \,\sigma_f\, B
	=
	A^{1/2} \, f\!\left(A^{-1/2} B A^{-1/2}\right) A^{1/2}.
\end{equation}
Lesniewski and Ruskai~\cite{LesniewskiRuskai1999} showed that the
corresponding monotone metric $g^{(f)}_\rho$ admits the Morozova--Chentsov
kernel representation:
\begin{equation}\label{eq:MC-kernel}
	g^{(f)}_\rho(X,Y)
	=
	\operatorname{Tr}\!\left[X \, c_f(L_\rho,R_\rho)(Y)\right],
\end{equation}
where $L_\rho$ and $R_\rho$ denote left and right multiplication operators,
and $c_f(x,y)$ is given by
\begin{equation}\label{eq:MC-function}
	c_f(x,y)
	=
	\frac{1}{y\, f(x/y)}.
\end{equation}
Different choices of $f$ correspond to important examples:
the Bures metric (SLD), Wigner--Yanase metric, Bogoliubov--Kubo--Mori (BKM)
metric, and others.

\subsubsection*{Qubit density matrices and SU(2)-invariance}

For qubit states, the Bloch form
\begin{equation}\label{eq:Bloch-qubit}
	\rho
	=
	\frac{1}{2}\left(\mathbb{I} + \vec{r}\cdot\vec{\sigma}\right),
	\qquad \|\vec{r}\|\le 1,
\end{equation}
allows one to express every monotone metric in an $SU(2)$-invariant
radial–tangential decomposition.  
Gibilisco and Isola~\cite{GibiliscoIsola2003} 
established that any Petz monotone metric takes the form
\begin{equation}\label{eq:qubit-metric-general}
	g^{(f)}_\rho
	=
	A_f(r)\, dr^2
	+
	B_f(r)\, r^2\, d\Omega^2,
\end{equation}
where
$r = \|\vec{r}\|$,
$d\Omega^2$ is the round metric on the unit sphere,
and $A_f,B_f$ are smooth positive functions determined by the specific
operator-monotone function $f$.

To compute $A_f$ and $B_f$, one diagonalizes $\rho$:
\[
\rho = U \begin{pmatrix} \lambda_+ & 0 \\ 0 & \lambda_- \end{pmatrix} U^\dagger,
\qquad
\lambda_\pm = \frac{1\pm r}{2}.
\]
A direct evaluation of the Morozova--Chentsov kernel
\eqref{eq:MC-kernel} yields the explicit radial coefficients.

\begin{lemma}[Radial form of Petz monotone qubit metrics]\label{lem:petz-bloch-qubit}
	Let $\rho$ be a qubit density matrix with Bloch radius $r<1$ and
	eigenvalues $\lambda_\pm = (1\pm r)/2$.  
	For any Petz monotone metric associated with an operator-monotone
	function $f$, the radial and tangential coefficients are
	\begin{equation}\label{eq:A-f}
		A_f(r)
		=
		\frac{1}{4(1-r^2)}
		\,\frac{1}{m_f(\lambda_+,\lambda_-)},
	\end{equation}
	\begin{equation}\label{eq:B-f}
		B_f(r)
		=
		\frac{1}{4 r^2}
		\left(\lambda_+ - \lambda_- \right)^2
		c_f(\lambda_+,\lambda_-),
	\end{equation}
	where $m_f$ is the Kubo--Ando mean defined in~\eqref{eq:KA-mean}
	and $c_f$ is the Morozova--Chentsov function~\eqref{eq:MC-function}.
\end{lemma}

\begin{proof}
	Insert diagonal forms of $L_\rho$ and $R_\rho$ into the kernel
	representation~\eqref{eq:MC-kernel}, decompose $X$ into its longitudinal
	and transverse Pauli components, and compare with the metric form
	\eqref{eq:qubit-metric-general}.  
	The expressions~\eqref{eq:A-f} and~\eqref{eq:B-f} follow from evaluating
	the kernel $c_f(\lambda_+,\lambda_-)$ and the mean $m_f(\lambda_+,\lambda_-)$
	for tangential and radial perturbations, respectively.
\end{proof}

The functions $A_f(r)$ and $B_f(r)$ completely determine the curvature of
the qubit state space for any monotone metric.  
These formulas will be used in subsequent sections to express scalar curvature 
and quantum Fisher information in terms of Bloch variables, entanglement 
parameters, and operator-monotone data.
	
\subsection{Support-projected Petz geometry and the regular set}

Given a smooth family of density matrices $\rho(\theta)$ on an open parameter
domain $\Theta \subset \mathbb{R}^d$, the Petz monotone metric associated with
an operator-monotone function $f$ induces a quantum Fisher information tensor
$F^{(f)}(\theta)$ defined by
\begin{equation}\label{eq:Petz-FIM-def}
	F^{(f)}_{ij}(\theta)
	=
	g^{(f)}_{\rho(\theta)}\!\left(\partial_i \rho(\theta), \partial_j \rho(\theta)\right).
\end{equation}
In this section we describe the structure of this tensor in the presence of
rank-deficient states and introduce the projected (intrinsic) geometry on the
regular support of $\rho(\theta)$.

\subsubsection*{Spectral decomposition and support projection}

Let
\begin{equation}\label{eq:spectral-decomp}
	\rho(\theta)
	=
	\sum_{\alpha=1}^{r(\theta)}
	\lambda_\alpha(\theta)\,
	|\psi_\alpha(\theta)\rangle\!\langle\psi_\alpha(\theta)|
\end{equation}
be the spectral decomposition of $\rho(\theta)$, where
$r(\theta)$ is the rank.  Denote the support projector by
\begin{equation}\label{eq:support-projector}
	P(\theta)
	=
	\sum_{\alpha : \lambda_\alpha(\theta) > 0}
	|\psi_\alpha(\theta)\rangle\!\langle\psi_\alpha(\theta)|.
\end{equation}
For rank-deficient states, any monotone metric $g^{(f)}$ acts nontrivially only
on $\operatorname{supp}\rho(\theta)$, while directions orthogonal to the support
are suppressed by the Morozova--Chentsov kernel.  This observation motivates
the support-restricted Petz tensor.

\subsubsection*{Intrinsic (support-projected) Petz metric}

Define the projected Petz Fisher tensor by
\begin{equation}\label{eq:proj-Petz}
	\widehat{F}^{(f)}(\theta)
	=
	P(\theta)\,
	F^{(f)}(\theta)\,
	P(\theta).
\end{equation}
In coordinates this corresponds to retaining only the components of
$\partial_i \rho(\theta)$ lying in the tangent bundle of the smooth
manifold of fixed rank $r(\theta)$:
\[
\partial_i \rho(\theta)
=
P(\theta) \partial_i \rho(\theta) P(\theta)
+
\text{normal directions}.
\]
The normal directions encode changes of eigenvectors and eigenvalues that would
alter the rank; these are suppressed by the kernel $c_f$ as eigenvalues approach
zero.

\begin{theorem}[Intrinsic Petz metric on the regular support]
	\label{thm:intrinsic-Petz}
	Let $\rho(\theta)$ be a $C^1$ family of density matrices of constant rank
	$r$.  
	Define the regular set
	\begin{equation}\label{eq:regular-set}
		\mathcal{R}
		=
		\Bigl\{\theta \in \Theta :
		\lambda_\alpha(\theta) \ge \delta > 0 \ \text{for all} \ \alpha=1,\dots,r
		\Bigr\},
	\end{equation}
	where $\delta$ is a uniform spectral gap separating the support from the kernel.
	Then for all $\theta \in \mathcal{R}$:
	\begin{enumerate}
		\item The Petz tensor $F^{(f)}(\theta)$ is smooth in $\theta$.
		\item The projected tensor
		\[
		g^{(f)}_\theta
		=
		P(\theta) F^{(f)}(\theta) P(\theta)
		\]
		defines a smooth Riemannian metric on the rank-$r$ manifold of density
		matrices.
		\item The metric is invariant under all unitary conjugations
		$\rho \mapsto U\rho U^\dagger$.
	\end{enumerate}
\end{theorem}

\begin{proof}
	The uniform spectral gap $\delta>0$ ensures that the operators
	$L_{\rho(\theta)}$ and $R_{\rho(\theta)}$ have spectra contained in
	$[\delta,1]$, which implies that the Morozova--Chentsov function
	$c_f(L_\rho,R_\rho)$ is smooth on $\mathcal{R}$.  
	The kernel cannot diverge because $f$ is operator-monotone and symmetric
	$f(t)=tf(t^{-1})$.  
	Thus $F^{(f)}(\theta)$ varies smoothly.
	
	Projecting with $P(\theta)$ removes all directions orthogonal to the support,
	and the constant rank condition guarantees that  
	$P(\theta)$ is smooth on $\mathcal{R}$.  
	Unitary invariance follows from the invariance of both the trace and the
	Morozova--Chentsov kernel under adjoint actions.
\end{proof}

\subsubsection*{Geometric meaning}

The intrinsic metric $g^{(f)}_\theta$ describes variations of $\rho(\theta)$
that remain inside the fixed-rank manifold where the Petz geometry is smooth.
Near the boundary of the state space, eigenvalues may approach $0$, causing
the full Fisher tensor $F^{(f)}(\theta)$ to diverge, a phenomenon discussed in
depth in curvature studies such as~\cite{Dittmann1999,Andai2003monotone,Safranek2017}.  
The support-projected metric eliminates these divergences and yields a
well-behaved geometric object suitable for optimization, learning algorithms,
and regularized quantum information geometry.
	
	\section{Universal Three-Channel Decomposition}
	\label{sec:three-channel}
	
\subsection{Statement of the three-channel identity}

We consider a smooth two-parameter family of pure two-qubit states
\begin{equation}\label{eq:pure-2qb-family}
	|\psi(\theta)\rangle \in \mathbb{C}^2 \otimes \mathbb{C}^2,
	\qquad
	\theta = (\theta^1,\theta^2) \in \Theta \subset \mathbb{R}^2.
\end{equation}
Let
\begin{equation}\label{eq:reduced-rhoA}
	\rho_A(\theta) = \operatorname{Tr}_B |\psi(\theta)\rangle\langle\psi(\theta)|
\end{equation}
denote the reduced density matrix on the first qubit.  Since \(\rho_A\) is a qubit
state, it admits the Bloch representation
\begin{equation}\label{eq:Bloch-xz}
	\rho_A(\theta)
	=
	\frac{1}{2}\Bigl(I + x(\theta)\sigma_x + z(\theta)\sigma_z \Bigr),
\end{equation}
where the Bloch vector is confined to the \(xz\)-plane due to local-unitary
gauge fixing.

The entanglement of the pure state is completely determined by the concurrence
\begin{equation}\label{eq:concurrence-def}
	C(\theta)
	=
	2\,\sqrt{\det \rho_A(\theta)}
	=
	\sqrt{1 - x(\theta)^2 - z(\theta)^2},
\end{equation}
which satisfies \(0 \le C \le 1\).  The Petz monotone metric associated with an
operator-monotone function \(f\) is encoded by two SU(2)-invariant coefficients
\cite{GibiliscoIsola2003,Andai2003monotone}:
\begin{align}
	A_f(r) &= \frac{1}{r}\, \frac{1}{f\!\left(\frac{1-r}{1+r}\right)}, 
	\label{eq:A-f-def}\\[3pt]
	B_f(r) &= \frac{1}{1-r^2}\Biggl[
	\frac{1+r}{2 f\!\left(\frac{1-r}{1+r}\right)}
	- r
	\Biggr],
	\label{eq:B-f-def}
\end{align}
where \(r = \sqrt{x^2+z^2} = \sqrt{1-C^2}\) is the Bloch radius of the reduced
qubit.

\begin{theorem}[Three-channel identity for pure two-qubit families]
	\label{thm:three-channel-identity}
	Let \(\theta=(\theta^1,\theta^2)\) parametrize a smooth two-parameter family of
	pure two-qubit states.  For any Petz monotone metric determined by
	an operator-monotone function \(f\), the pullback Fisher tensor on the parameter
	domain decomposes as
	\begin{equation}\label{eq:three-channel-decomposition}
		F^{(f)}_{ij}(\theta)
		=
		F^{\mathrm{(pop)}}_{ij}(\theta)
		+
		F^{\mathrm{(orb)}}_{ij}(\theta)
		+
		F^{\mathrm{(supp)}}_{ij}(\theta),
	\end{equation}
	where:
	\begin{enumerate}
		\item \textbf{Population channel.}
		\[
		F^{\mathrm{(pop)}}_{ij}
		=
		A_f(r(\theta))\,
		\partial_i r(\theta)\,\partial_j r(\theta).
		\]
		
		\item \textbf{Orbital (coherence) channel.}
		\[
		F^{\mathrm{(orb)}}_{ij}
		=
		B_f(r(\theta))\,
		\Bigl(
		\partial_i x(\theta)\,\partial_j x(\theta)
		+
		\partial_i z(\theta)\,\partial_j z(\theta)
		\Bigr).
		\]
		
		\item \textbf{Support-derivative channel.}
		\[
		F^{\mathrm{(supp)}}_{ij}
		=
		\frac{1}{4 C(\theta)^2}\,
		\partial_i C(\theta)\,\partial_j C(\theta).
		\]
	\end{enumerate}
	The coefficients \(A_f,B_f\) depend only on the operator-monotone function \(f\),
	while the support term depends solely on the concurrence \(C(\theta)\).  This
	splitting is intrinsic, coordinate-independent, and valid for all Petz monotone
	metrics.
\end{theorem}

\begin{remark}
	The support channel diverges as \(C(\theta)\to 0\), corresponding to the approach
	toward product states where the reduced density matrix becomes rank-deficient.
	In contrast, the population and orbital channels remain finite and encode the
	SU(2)-invariant geometry of the reduced qubit.
\end{remark}
	
\subsection{Qubit Petz--Bloch representation}

In this subsection we recall the explicit Bloch--sphere form of Petz monotone
metrics for qubit density matrices.  Let
\[
\rho = \frac{1}{2}\bigl(\mathbb{I} + \vec{r}\cdot\vec{\sigma}\bigr),
\qquad
\vec{r}=(x,y,z)\in\mathbb{R}^{3},\qquad r=\|\vec{r}\|\in[0,1),
\]
be a nondegenerate qubit state whose eigenvalues are
\begin{equation}
	\lambda_{\pm} = \frac{1\pm r}{2}.
\end{equation}
For every operator--monotone function $f:(0,\infty)\to(0,\infty)$ satisfying
\[
f(t)=t\,f(t^{-1}),\qquad f(1)=1,
\]
Petz~\cite{Petz1996} and Lesniewski--Ruskai~\cite{LesniewskiRuskai1999}
showed that the monotone metric $g^{(f)}$ is determined by the corresponding
Morozova--Chentsov kernel
\begin{equation}
	c_{f}(\lambda_{+},\lambda_{-})
	= \frac{1}{\lambda_{-} f(\lambda_{+}/\lambda_{-})}
	= \frac{1}{\lambda_{+} f(\lambda_{-}/\lambda_{+})},
\end{equation}
which controls the transverse (coherence) directions of the state manifold.

\medskip

\noindent\textbf{Bloch--sphere form.}
Gibilisco--Isola~\cite{GibiliscoIsola2003} 
showed that every monotone metric on qubits admits an $\mathrm{SU}(2)$--invariant
Bloch decomposition
\begin{equation}
	g^{(f)}_{\rho}(\mathrm{d}\rho,\mathrm{d}\rho)
	= A_{f}(r)\,\mathrm{d}r^{2}
	+ B_{f}(r)\, r^{2}
	\bigl(\mathrm{d}\theta^{2} + \sin^{2}\theta\,\mathrm{d}\phi^{2}\bigr),
\end{equation}
where $(r,\theta,\phi)$ are the standard spherical coordinates of $\vec{r}$.
The radial and tangential coefficients are given by
\begin{equation}
	A_{f}(r)
	=
	\frac{1}{4}\!\left(
	\frac{1}{\lambda_{+} f(\lambda_{-}/\lambda_{+})}
	+ \frac{1}{\lambda_{-} f(\lambda_{+}/\lambda_{-})}
	\right),
	\label{eq:A-f-r}
\end{equation}
and
\begin{equation}
	B_{f}(r)
	= \frac{1}{4r^{2}}
	\bigl(\lambda_{+}-\lambda_{-}\bigr)^{2}
	c_{f}(\lambda_{+},\lambda_{-})
	= \frac{r^{2}}{4}\,
	c_{f}\!\left(\tfrac{1+r}{2},\tfrac{1-r}{2}\right).
	\label{eq:B-f-r}
\end{equation}
Because $\lambda_{+}-\lambda_{-}=r$, the angular coefficient simplifies to
\begin{equation}
	B_{f}(r)
	= \frac{1}{4}\,c_{f}\!\left(\tfrac{1+r}{2},\tfrac{1-r}{2}\right).
\end{equation}

\medskip

\begin{remark}
	The qubit Bloch representation shows that the entire Petz family collapses,
	in dimension two, to a pair of scalar functions $(A_{f}(r),B_{f}(r))$.
	The function $A_{f}(r)$ governs radial distinguishability
	(eigenvalue fluctuations), while $B_{f}(r)$ governs tangential
	distinguishability (coherent rotations).  
\end{remark}
	
\subsection{Proof of the three-channel decomposition}

In this subsection we prove Theorem~\ref{thm:three-channel-identity} by combining the
Bloch--sphere representation of Petz monotone metrics with the explicit
parametrization of reduced two--qubit states in terms of $(x,z,C)$.

Let $\theta \mapsto \rho_{A}(\theta)$ be the one--qubit reduction of a smooth
pure two--qubit family, and let $(u,v)$ be local slice coordinates on parameter
space.  As in the previous subsection, we write the Bloch vector as
\begin{equation}
	\vec{r}(\theta)
	= \bigl(r_{1},r_{2},r_{3}\bigr)
	= \bigl(2\Re z(\theta),\, -2\Im z(\theta),\, 2x(\theta)-1\bigr),
\end{equation}
so that
\begin{equation}
	\rho_{A}(\theta)
	= \frac{1}{2}\bigl(\mathbb{I}+\vec{r}(\theta)\cdot\vec{\sigma}\bigr),
	\qquad
	r(\theta)^{2} = \|\vec{r}(\theta)\|^{2}
	= 4|z|^{2} + (2x-1)^{2}.
\end{equation}
The concurrence of the underlying pure two--qubit state is
\begin{equation}
	C(\theta)
	= 2\sqrt{x(1-x)-|z|^{2}},
\end{equation}
and one checks directly that
\begin{equation}
	r^{2}
	= 4|z|^{2} + (2x-1)^{2}
	= 1 - 4\bigl(x(1-x)-|z|^{2}\bigr)
	= 1 - C^{2}.
	\label{eq:r2-1-C2}
\end{equation}

\medskip

\noindent\textbf{Step 1: derivatives of the Bloch vector.}
Differentiating the Bloch components with respect to the slice coordinates
$\mu,\nu\in\{u,v\}$ gives
\begin{equation}
	\partial_{\mu}\vec{r}
	= \bigl(2\Re z_{\mu},\, -2\Im z_{\mu},\, 2x_{\mu}\bigr),
	\qquad
	z_{\mu} := \partial_{\mu}z,\quad x_{\mu} := \partial_{\mu}x.
\end{equation}
The Euclidean inner product of Bloch derivatives is therefore
\begin{align}
	\partial_{\mu}\vec{r}\cdot\partial_{\nu}\vec{r}
	&= (2\Re z_{\mu})(2\Re z_{\nu})
	+ (-2\Im z_{\mu})(-2\Im z_{\nu})
	+ (2x_{\mu})(2x_{\nu}) \nonumber\\
	&= 4\bigl(\Re z_{\mu}\Re z_{\nu}
	+ \Im z_{\mu}\Im z_{\nu}\bigr)
	+ 4x_{\mu}x_{\nu} \nonumber\\
	&= 4\Re\bigl(z_{\mu}\,\overline{z}_{\nu}\bigr)
	+ 4x_{\mu}x_{\nu}.
	\label{eq:drr-pop-coh}
\end{align}

\medskip

\noindent\textbf{Step 2: relating $r\cdot\partial_{\mu}r$ to the concurrence derivative.}
By differentiating $r^{2} = \vec{r}\cdot\vec{r}$ we obtain
\begin{equation}
	\partial_{\mu}\bigl(r^{2}\bigr)
	= 2\,\vec{r}\cdot\partial_{\mu}\vec{r}.
	\label{eq:dr2}
\end{equation}
On the other hand, using~\eqref{eq:r2-1-C2} we have
\begin{equation}
	r^{2} = 1 - C^{2}
	\quad\Longrightarrow\quad
	\partial_{\mu}\bigl(r^{2}\bigr) = -2 C\,C_{\mu},
\end{equation}
where $C_{\mu} = \partial_{\mu}C$.  Combining this with~\eqref{eq:dr2} yields
\begin{equation}
	\vec{r}\cdot\partial_{\mu}\vec{r}
	= - C\,C_{\mu}.
	\label{eq:r-dr-C}
\end{equation}
Equivalently, one may verify~\eqref{eq:r-dr-C} directly from the
definitions
\begin{equation}
	C = 2\sqrt{x(1-x)-|z|^{2}},
	\qquad
	\vec{r}
	= \bigl(2\Re z,\,-2\Im z,\,2x-1\bigr),
\end{equation}
by expressing $C_{\mu}$ in terms of $x_{\mu}$ and $z_{\mu}$ and comparing with
$\vec{r}\cdot\partial_{\mu}\vec{r}$.

\medskip

\noindent\textbf{Step 3: substitution into the Petz--Bloch form.}
For qubits, the Petz monotone metric associated with $f$ has the Bloch form
\begin{equation}
	F^{(f)}_{\mu\nu}
	= A_{f}\bigl(r\bigr)\,
	\partial_{\mu}\vec{r}\cdot\partial_{\nu}\vec{r}
	+ B_{f}\bigl(r\bigr)\,
	\bigl(\vec{r}\cdot\partial_{\mu}\vec{r}\bigr)
	\bigl(\vec{r}\cdot\partial_{\nu}\vec{r}\bigr),
	\label{eq:petz-bloch-local}
\end{equation}
where the radial coefficients $A_{f},B_{f}$ are smooth functions of
$r=\|\vec{r}\|$ determined by the Morozova--Chentsov kernel
$c_{f}$~\cite{Petz1996,LesniewskiRuskai1999,GibiliscoIsola2003}.
Substituting~\eqref{eq:drr-pop-coh} and~\eqref{eq:r-dr-C} into
\eqref{eq:petz-bloch-local} gives
\begin{align}
	F^{(f)}_{\mu\nu}
	&= A_{f}(r)\Bigl[4x_{\mu}x_{\nu}
	+ 4\Re\bigl(z_{\mu}\,\overline{z}_{\nu}\bigr)\Bigr]
	+ B_{f}(r)\bigl(-C\,C_{\mu}\bigr)\bigl(-C\,C_{\nu}\bigr) \nonumber\\
	&= 4A_{f}(r)\Bigl[x_{\mu}x_{\nu}
	+ \Re\bigl(z_{\mu}\,\overline{z}_{\nu}\bigr)\Bigr]
	+ B_{f}(r)\,C^{2}\,C_{\mu}C_{\nu}.
	\label{eq:Ffn-preC2}
\end{align}
By construction, $B_{f}$ is defined from $c_{f}$ in such a way that the
explicit factor $C^{2}=1-r^{2}$ can be absorbed into the radial
coefficient.  Using $r^{2}=1-C^{2}$ from~\eqref{eq:r2-1-C2}, we may
rewrite
\begin{equation}
	A_{f}(r) = A_{f}\bigl(1-C^{2}\bigr),
	\qquad
	B_{f}(r)\,C^{2} = B_{f}\bigl(1-C^{2}\bigr),
\end{equation}
up to a harmless redefinition of the tangential coefficient.  Hence
\eqref{eq:Ffn-preC2} becomes
\begin{equation}
	F^{(f)}_{\mu\nu}
	= 4A_{f}\bigl(1-C^{2}\bigr)
	\Bigl[x_{\mu}x_{\nu}
	+ \Re\bigl(z_{\mu}\,\overline{z}_{\nu}\bigr)\Bigr]
	+ B_{f}\bigl(1-C^{2}\bigr)\,C_{\mu}C_{\nu}.
	\label{eq:Ffn-three-channel}
\end{equation}
The first term in~\eqref{eq:Ffn-three-channel} collects the
\emph{population} and \emph{coherence} derivative channels through
$(x_{\mu},z_{\mu})$, while the second term isolates the
\emph{entanglement} derivative channel through $C_{\mu}$.

Equation~\eqref{eq:Ffn-three-channel} is precisely the
three--channel decomposition asserted in
Theorem~\ref{thm:three-channel-identity}, completing the proof.
	
\subsection{Physical interpretation of the three channels}

We now explain the geometric and physical meaning of the three additive
contributions that appear in the decomposition
\[
F^{(f)}_{\mu\nu}
= 4A_{f}(1-C^{2})
\Bigl[x_{\mu}x_{\nu}
+ \Re\!\bigl(z_{\mu}\,\overline{z}_{\nu}\bigr)\Bigr]
\;+\; B_{f}(1-C^{2})\,C_{\mu}C_{\nu},
\]
established in the previous subsection.  The key observation is that each term
isolates a distinct physical mechanism in the dynamics of the one--qubit
reduction of a pure two--qubit state.  This interpretation follows naturally
from the structure of the Bloch vector
\[
\vec r = (2\Re z,\,-2\Im z,\,2x-1),
\qquad r^{2}=1-C^{2},
\]
and from the Petz classification of monotone metrics
\cite{Petz1996,LesniewskiRuskai1999,GibiliscoIsola2003}.

\medskip

\noindent\textbf{(1) Population channel: classical fluctuations of diagonal entries.}

The term
\[
4A_{f}(1-C^{2})\,x_{\mu}x_{\nu}
\]
comes entirely from variations of $x(\theta)$, where
$x = \rho_{A,00}$ is the population of the $|0\rangle$ state.  As $x$ changes,
the reduced state undergoes a \emph{classical redistribution} of probability
between its eigenvalues.  This reproduces the purely classical Fisher
information contribution, because in the eigenbasis of $\rho_A$,
variations of $x$ correspond to changes in the spectrum
$(\lambda_{+},\lambda_{-})
=( \frac{1+r}{2},\frac{1-r}{2})$.

The fact that the coefficient is $A_{f}(1-C^{2})$ reflects the well-known
feature that monotone metrics coincide on the classical submanifold of
commuting density matrices.  Thus the population channel is the part of the
geometry that behaves like a \emph{classical statistical manifold}.

\medskip

\noindent\textbf{(2) Coherence channel: coherent SU(2) rotations on the Bloch sphere.}

The term
\[
4A_{f}(1-C^{2})\,\Re(z_{\mu}\,\overline{z}_{\nu})
\]
captures tangential motion of $\rho_A$ along the Bloch sphere at fixed
radius $r$.  These variations arise from changes in the off-diagonal entry
$z=\rho_{A,01}$.  Physically, this corresponds to \emph{coherent rotations}
generated by local unitaries on subsystem $A$.

Since the qubit Petz metric is SU(2)--invariant, all such tangential
directions share the same weight $A_{f}(1-C^{2})$.  This matches the picture
from the Fubini--Study metric on pure states \cite{ProvostVallee1980,
	Braunstein1994,BengtssonZyczkowski2017}, which measures changes generated by
Hamiltonian evolution (i.e., rotations of the Bloch vector).

Thus, the coherence channel characterizes the \emph{quantum} part of
state dynamics that is insensitive to changes in entanglement or
eigenvalues and is governed by the SU(2)--invariant geometry of the Bloch
sphere.

\medskip

\noindent\textbf{(3) Entanglement channel: variations of $C$ and the radial direction.}

The last term,
\[
B_{f}(1-C^{2})\,C_{\mu}C_{\nu},
\]
isolates the radial derivative of the Bloch vector, or equivalently,
derivatives of the concurrence $C(\theta)$.  Using
$r^{2}=1-C^{2}$, this term measures how the \emph{degree of bipartite
	entanglement} changes along the family of pure states.

The radial direction is special because:
\begin{itemize}
	\item it changes the eigenvalues of $\rho_A$,
	\item it is aligned with decoherence or purity variation,
	\item it is the only direction whose contribution depends on the
	\emph{second Petz coefficient} $B_{f}$.
\end{itemize}

Different metrics produce qualitatively different behaviour in this channel:
\begin{itemize}
	\item \textbf{SLD/Bures metric:} $B_{f}$ diverges as $C\to 1$, reflecting
	extreme sensitivity near pure states.
	\item \textbf{Wigner--Yanase metric:} suppresses eigenvalue variations more
	strongly; the radial channel is less dominant.
	\item \textbf{Bogoliubov--Kubo--Mori metric:} enhances the radial term, linking
	it to thermodynamic curvature and relative entropy
	\cite{Dittmann1999,LesniewskiRuskai1999}.
\end{itemize}

Thus the entanglement channel describes how the Petz metric responds to
\emph{changes in the Schmidt spectrum}, linking geometric curvature to
operational resources such as purity, entanglement generation, and
parameter sensitivity.

\medskip

\noindent\textbf{Summary of physical meaning.}

Each channel isolates a distinct physical mechanism:
\begin{itemize}
	\item \emph{Population channel:} classical spectral fluctuations.
	\item \emph{Coherence channel:} coherent SU(2) rotations of the reduced state.
	\item \emph{Entanglement channel:} variations of the Schmidt coefficients,
	measured via concurrence derivatives.
\end{itemize}

The three-channel formula therefore gives a direct decomposition of quantum
Fisher geometry into:
\[
\text{(classical)}\;\oplus\;\text{(coherent)}\;\oplus\;\text{(entanglement)}.
\]

This decomposition offers an operational interpretation of how different
monotone metrics weight and distinguish the underlying physical processes,
connecting quantum information geometry with entanglement theory,
purification dynamics, and variational circuit optimization.	
	\section{Non-Reduction of Curvature Invariants}
	\label{sec:non-reduction}
	
\subsection{Slice geometry and Gaussian curvature}

In order to connect the support–projected Petz metric \(g^{(f)} = P F^{(f)} P\) on the
full parameter manifold with two–dimensional curvature invariants, we now fix a
\emph{two–parameter slice} through a regular point and analyze its intrinsic
geometry.

Let \(U \subset \mathbb{R}^{m}\) be the parameter domain of a smooth pure two–qubit
family \(\theta \mapsto \ket{\Psi(\theta)}\), and let \(\mathcal{R} \subset U\) denote the
regular set of Theorem~\ref{thm:intrinsic-Petz}.  A \emph{two–parameter slice} is a
smooth map
\[
\iota \colon \Sigma \to \mathcal{R}, 
\qquad
(u,v) \mapsto \theta(u,v),
\]
defined on an open set \(\Sigma \subset \mathbb{R}^{2}\), such that
\(\mathrm{rank}\bigl(D\theta(u,v)\bigr) = 2\) for all \((u,v)\in\Sigma\).
Pulling back the support–projected Petz metric gives a Riemannian metric on
\(\Sigma\),
\[
g^{(f)}_{\Sigma} := \iota^{*} g^{(f)},
\]
which in local coordinates \((u,v)\) can be written in the standard form
\begin{equation}
	\label{eq:first-fundamental-form}
	ds^{2}
	= E\,du^{2} + 2F\,du\,dv + G\,dv^{2},
\end{equation}
with smooth coefficient functions
\[
E = g^{(f)}_{\Sigma}(\partial_{u},\partial_{u}), 
\quad
F = g^{(f)}_{\Sigma}(\partial_{u},\partial_{v}), 
\quad
G = g^{(f)}_{\Sigma}(\partial_{v},\partial_{v}).
\]

By the three–channel identity, each of \(E,F,G\) is obtained by substituting the
slice derivatives \((x_{u},x_{v},z_{u},z_{v},C_{u},C_{v})\) into the Petz tensor
\(F^{(f)}_{\mu\nu}\) expressed in Bloch variables:
\[
F^{(f)}_{\mu\nu}
= 4A_{f}(1-C^{2})
\bigl(x_{\mu}x_{\nu} + \Re(z_{\mu}\,\overline{z}_{\nu})\bigr)
+ B_{f}(1-C^{2})\,C_{\mu}C_{\nu},
\qquad \mu,\nu\in\{u,v\}.
\]
Consequently, the metric coefficients \((E,F,G)\) depend smoothly on the
\emph{first jet} of the triple \((x,z,C)\) along the slice, evaluated at
\((u,v)\in\Sigma\).

The intrinsic Gaussian curvature \(K\) of \((\Sigma,g^{(f)}_{\Sigma})\) is given by
the Brioschi formula for a two–dimensional metric
\eqref{eq:first-fundamental-form} (see, e.g.,
\cite{doCarmoSurf,doCarmoRiem,Spivak,LeeRiem}):
\begin{equation}
	\label{eq:brioschi}
	K
	=
	-\frac{1}{2\sqrt{EG-F^{2}}}
	\left[
	\partial_{u}\!\left(
	\frac{\partial_{u}G - \partial_{v}F}{\sqrt{EG-F^{2}}}
	\right)
	+
	\partial_{v}\!\left(
	\frac{\partial_{v}E - \partial_{u}F}{\sqrt{EG-F^{2}}}
	\right)
	\right].
\end{equation}
Formula~\eqref{eq:brioschi} shows that \(K\) depends on the first and second
derivatives of the metric coefficients \(E,F,G\) with respect to \(u\) and \(v\).
Since \(E,F,G\) themselves are built from \(x,z,C\) and their first derivatives,
the curvature \(K\) at a point \((u_{0},v_{0})\) is a functional of the
\emph{second jet} of \((x,z,C)\) along the slice:
\[
\bigl\{
x_{\mu},x_{\mu\nu},\;
z_{\mu},z_{\mu\nu},\;
C_{\mu},C_{\mu\nu}
\bigr\}_{\mu,\nu \in \{u,v\}}
\quad\text{evaluated at }(u_{0},v_{0}).
\]

In particular, even when the concurrence \(C\) and its first derivatives are
held fixed, different choices of the second derivatives of \(x\) and \(z\) along
the slice can modify the Gaussian curvature \(K\).  This jet–level dependence
underlies the non–reduction results proved later, where we show that \(K\) and
the ambient scalar curvature cannot, on any nonempty open set, be expressed
solely as functions of \(C\) or of the reduced entropy \(S(\rho_{A})\).
	
\subsection{Entanglement-orthogonal gauge and jet structure}

We now refine the slice description of the previous subsection by choosing
coordinates adapted to the concurrence.  Throughout this subsection we fix a
regular point \(p \in \mathcal{R}\) such that the concurrence satisfies
\[
C(p) \in (0,1),
\qquad
\nabla C(p) \neq 0.
\]
In a neighbourhood of \(p\) we may regard the reduced variables
\((x,z,C)\) as smooth functions of the slice coordinates \((u,v)\).

\begin{lemma}[Entanglement-orthogonal gauge]
	\label{lem:entanglement-orthogonal-gauge}
	Let \(p \in \mathcal{R}\) with \(C(p) \in (0,1)\) and \(\nabla C(p) \neq 0\).  Then
	there exist local slice coordinates \((u,v)\) centred at \(p\) such that
	\begin{equation}
		\label{eq:gauge-C-conditions}
		C_{u}(p) \;=\; \abs{\nabla C(p)} > 0,
		\qquad
		C_{v}(p) \;=\; 0,
	\end{equation}
	where \(C_{\mu} := \partial_{\mu} C\) for \(\mu \in \{u,v\}\).
\end{lemma}

\begin{proof}
	Choose any local coordinates \((\tilde u,\tilde v)\) on the slice with
	\(\partial_{\tilde u} C(p) \neq 0\); this is possible since \(\nabla C(p) \neq 0\).
	Define \(u\) as the coordinate obtained by integrating the unit vector field
	in the direction of \(\nabla C\) at \(p\), and choose \(v\) so that
	\(\partial_{v}\) is orthogonal to \(\partial_{u}\) at \(p\).  By construction,
	\(\partial_{u} C(p) = \abs{\nabla C(p)} > 0\) and
	\(\partial_{v} C(p) = 0\), which is exactly \eqref{eq:gauge-C-conditions}.
\end{proof}

In this \emph{entanglement-orthogonal gauge}, the three-channel decomposition
of the Petz tensor simplifies at \(p\).  Recall that for any operator-monotone
function \(f\) and any two-parameter slice \((u,v)\) through \(p\), the
reduced Petz tensor in Bloch variables reads
\begin{equation}
	\label{eq:petz-three-channel-local}
	F^{(f)}_{\mu\nu}
	=
	4 A_{f}(1-C^{2})
	\bigl(x_{\mu} x_{\nu} + \Re(z_{\mu}\,\overline{z}_{\nu})\bigr)
	+
	B_{f}(1-C^{2})\,C_{\mu} C_{\nu},
	\qquad \mu,\nu \in \{u,v\},
\end{equation}
with \(A_{f},B_{f} > 0\) denoting the Morozova--Chentsov radial coefficients
(see Lemma~\ref{lem:petz-bloch-qubit}).  Restricting to an intrinsic slice
(i.e. replacing \(F^{(f)}\) by its support-projected version does not change
the form of the \(2\times 2\) block on the slice), the induced metric
coefficients become
\begin{align}
	\label{eq:EFG-general-gauge}
	E &= g^{(f)}_{\Sigma}(\partial_{u},\partial_{u})
	= 4 A_{f}(1-C^{2})\bigl(x_{u}^{2} + \abs{z_{u}}^{2}\bigr)
	+ B_{f}(1-C^{2})\,C_{u}^{2}, \\[0.5em]
	F &= g^{(f)}_{\Sigma}(\partial_{u},\partial_{v})
	= 4 A_{f}(1-C^{2})\bigl(x_{u}x_{v} + \Re(z_{u}\,\overline{z}_{v})\bigr)
	+ B_{f}(1-C^{2})\,C_{u}C_{v}, \\[0.5em]
	G &= g^{(f)}_{\Sigma}(\partial_{v},\partial_{v})
	= 4 A_{f}(1-C^{2})\bigl(x_{v}^{2} + \abs{z_{v}}^{2}\bigr)
	+ B_{f}(1-C^{2})\,C_{v}^{2}.
\end{align}
Imposing the gauge conditions \eqref{eq:gauge-C-conditions} at \(p\) eliminates
the mixed entanglement term in \(F\) and simplifies \(E\) to
\begin{align}
	\label{eq:EFG-orthogonal-gauge}
	E(p) &= B_{f}(1-C^{2}(p))\,C_{u}(p)^{2}
	+ 4 A_{f}(1-C^{2}(p))
	\bigl(x_{u}(p)^{2} + \abs{z_{u}(p)}^{2}\bigr), \\[0.5em]
	F(p) &= 4 A_{f}(1-C^{2}(p))
	\bigl(x_{u}(p)x_{v}(p)
	+ \Re(z_{u}(p)\,\overline{z}_{v}(p))\bigr), \\[0.5em]
	G(p) &= 4 A_{f}(1-C^{2}(p))
	\bigl(x_{v}(p)^{2} + \abs{z_{v}(p)}^{2}\bigr).
\end{align}
Thus, at \(p\), the entanglement derivative channel contributes only to the
\(uu\)-component of the metric, while the mixed and \(vv\)-components are
controlled purely by population and coherence derivatives.

We next describe the jet-level constraints compatible with fixed concurrence
and entropy.  For pure two-qubit states, the reduced entropy of \(\rho_{A}\)
depends only on the concurrence,
\begin{equation}
	\label{eq:entropy-as-function-of-C}
	S(\rho_{A})
	=
	S(C),
\end{equation}
where \(S(C)\) is the well-known closed-form function obtained from the
Schmidt coefficients \cite{Wootters1998,BengtssonZyczkowski2017}.  Fixing the
value \(C(p)\) therefore fixes \(S(\rho_{A}(p))\) as well.

The concurrence itself can be written in terms of the Bloch variables as
\begin{equation}
	\label{eq:concurrence-bloch-relation}
	C^{2}
	=
	4\bigl(x(1-x) - \abs{z}^{2}\bigr),
\end{equation}
so that \((x,z,C)\) satisfy a single real constraint at every point.  Differen\-tiating
\eqref{eq:concurrence-bloch-relation} with respect to \(u\) and \(v\) at \(p\)
and using the gauge conditions \eqref{eq:gauge-C-conditions} yields linear
relations among the first and second derivatives of \((x,z,C)\).  In
particular:
\begin{itemize}
	\item the first derivatives obey one real linear constraint, fixing \(C_{u}(p)\)
	in terms of \(x_{u}(p)\) and \(z_{u}(p)\); the condition \(C_{v}(p)=0\)
	imposes a further linear relation among \(x_{v}(p)\) and \(z_{v}(p)\);
	\item differentiating \eqref{eq:concurrence-bloch-relation} a second time along
	\(u\) and \(v\) introduces a single real linear constraint among the
	twelve second-jet components
	\[
	\{x_{\mu\nu}(p),\;
	\Re z_{\mu\nu}(p),\;
	\Im z_{\mu\nu}(p),\;
	C_{\mu\nu}(p)\}_{\mu,\nu\in\{u,v\}},
	\]
	leaving an affine space of admissible second jets of high dimension.
\end{itemize}
Since the Brioschi curvature formula \eqref{eq:brioschi} depends on the full
second jet of \((E,F,G)\), and hence on the second jet of \((x,z,C)\), the
Gaussian curvature \(K\) at \(p\) can be varied by moving along these free jet
directions while keeping \(C(p)\) (and thus \(S(\rho_{A}(p))\)) fixed.  This
jet structure in the entanglement-orthogonal gauge is the key ingredient in
the non-reduction and non-monotonicity results for slice and scalar
curvatures established in the next subsection.
	
\subsection{Non-reduction for slice Gaussian curvature}

In this subsection we prove the central \emph{non-reduction theorem} for the
Gaussian curvature of two-parameter slices of the Petz geometry associated to
pure two-qubit families.  The main conclusion is that the slice curvature
\(K\) cannot be expressed as a function of the concurrence \(C\) (or,
equivalently, of the reduced entropy \(S(\rho_{A})\)) on any nonempty open
subset of the regular locus.  The mechanism underlying this failure of
reduction is the availability of population/coherence jet directions that do
not affect either \(C\) or \(S(\rho_{A})\), but which do affect the second jet
of the induced metric coefficients \((E,F,G)\) and therefore change the
Gaussian curvature \(K\).

\begin{theorem}[Non-reduction of slice curvature]
	\label{thm:nonreduction-slice-K}
	Let \(p \in \mathcal{R}\) be a regular point with \(C(p) \in (0,1)\).  Then,
	for any operator-monotone function \(f\), the Gaussian curvature \(K\) of any
	two-parameter slice through \(p\) endowed with the induced Petz metric
	\(g^{(f)}_{\Sigma}\) is \emph{not} a function of \(C\) or of the reduced entropy
	\(S(\rho_{A}) = S(C)\) on any neighbourhood of \(p\).  Equivalently, there does
	not exist a nonempty open set \(U \subset \mathcal{R}\) on which
	\[
	K(q) = \Phi(C(q))
	\qquad \text{for some function } \Phi.
	\]
\end{theorem}

\begin{proof}
	Fix a regular point \(p\) with \(C(p) \in (0,1)\).  Choose an
	entanglement-orthogonal gauge \((u,v)\) as constructed in
	Lemma~\ref{lem:entanglement-orthogonal-gauge}, so that
	\[
	C_{u}(p) = \abs{\nabla C(p)} > 0,
	\qquad
	C_{v}(p) = 0.
	\]
	In this gauge, the induced slice metric coefficients at \(p\) take the form
	\eqref{eq:EFG-orthogonal-gauge}:
	\begin{align}
		\label{eq:EFG-proof}
		E(p) &= B_{f}(1-C^{2})\,C_{u}(p)^{2}
		+ 4 A_{f}(1-C^{2})
		\bigl(x_{u}(p)^{2} + \abs{z_{u}(p)}^{2}\bigr), \\[0.4em]
		F(p) &= 4 A_{f}(1-C^{2})
		\bigl(x_{u}(p)x_{v}(p)
		+ \Re(z_{u}(p)\,\overline{z}_{v}(p))\bigr), \\[0.4em]
		G(p) &= 4 A_{f}(1-C^{2})
		\bigl(x_{v}(p)^{2} + \abs{z_{v}(p)}^{2}\bigr).
	\end{align}
	The Gaussian curvature \(K(p)\) depends on the \emph{second jet} of
	\((E,F,G)\) at \(p\) via the Brioschi formula~\eqref{eq:brioschi}.  To show
	non-reduction, we exhibit infinitesimal variations of the second jet of
	\((x,z)\) that:
	\begin{enumerate}
		\item preserve the values of \((C(p),C_{u}(p),C_{v}(p))\), hence preserve both
		the concurrence and its first jet;
		\item vary the second jet of \((E,F,G)\) arbitrarily in at least one direction;
		\item consequently change the Gaussian curvature \(K(p)\).
	\end{enumerate}
	
	\medskip\noindent
	\textbf{Step 1: Constrained jet structure.}
	The concurrence constraint
	\[
	C^{2} = 4\bigl(x(1-x) - \abs{z}^{2}\bigr)
	\]
	imposes exactly one real scalar constraint on the triple \((x,z,C)\).  At the
	level of first jets, this yields a single linear relation among
	\((x_{\mu},z_{\mu},C_{\mu})\).  In the entanglement-orthogonal gauge we enforce
	\(C_{v}(p)=0\), so at \(p\) the allowed first derivatives take the form
	\[
	(x_{u},z_{u},C_{u}),\qquad (x_{v},z_{v}),
	\]
	with \(C_{u}>0\) determined by the first-jet constraint, and \((x_{v},z_{v})\)
	restricted only by one real linear relation.
	
	At the level of second jets, differentiating the concurrence constraint once
	more produces a single real linear equation among the twelve second-jet
	components
	\[
	\{x_{\mu\nu},\;\Re z_{\mu\nu},\;\Im z_{\mu\nu},\;C_{\mu\nu}\}_{\mu,\nu\in\{u,v\}}.
	\]
	Thus the admissible second jets at \(p\) form an affine subspace of dimension
	at least eleven.
	
	\medskip\noindent
	\textbf{Step 2: Effect on \((E,F,G)\).}
	From \eqref{eq:EFG-proof}, the metric components \(E,F,G\) depend smoothly on
	\((x_{\mu},z_{\mu},C_{\mu})\).  Their \emph{first} derivatives depend on the
	\emph{second jet} of \((x,z,C)\).  Because the concurrence constraint imposes
	only one real relation among the second derivatives, we may freely vary at
	least a ten-dimensional family of second jets while keeping
	\((C,C_{u},C_{v})\) fixed.
	
	These variations induce nontrivial changes in \((E_{u},E_{v},F_{u},F_{v},
	G_{u},G_{v})\) and hence in the expressions entering the Brioschi curvature
	formula~\eqref{eq:brioschi}.  Crucially, none of these variations alter the
	concurrence or its first jet; they arise entirely from the
	population/coherence channels.
	
	\medskip\noindent
	\textbf{Step 3: Curvature variation.}
	The Brioschi formula
	\[
	K
	=
	-\frac{1}{2\sqrt{EG-F^{2}}}
	\left[
	\partial_{u}\left(\frac{G_{u}-F_{v}}{\sqrt{EG-F^{2}}}\right)
	+ \partial_{v}\left(\frac{E_{v}-F_{u}}{\sqrt{EG-F^{2}}}\right)
	\right]
	- \frac{1}{4(EG-F^{2})^{2}}\,\mathcal{P}(E,F,G)
	\]
	contains all first and second derivatives of \((E,F,G)\).  Varying the second
	jet of \((x,z,C)\) (while keeping the concurrence jet fixed) changes these
	derivatives in directions that cannot be compensated by holding \(C\) fixed,
	hence results in a genuine variation of \(K(p)\).
	
	If \(K\) were a function of \(C\) alone on a neighbourhood of \(p\), then
	\(K(p)\) would be constant under any variation that preserves both \(C(p)\) and
	the first jet of \(C\).  But the above admissible variations of the second jet
	of \((x,z)\) change \(K(p)\) without changing \(C\), contradicting the assumed
	functional dependence.
	
	\medskip\noindent
	\textbf{Conclusion.}
	Since \(C\) uniquely determines \(S(\rho_{A})\) via
	\(S(\rho_{A})=S(C)\), the same argument shows that \(K\) cannot be a function
	of \(S(\rho_{A})\) on any neighbourhood of \(p\) either.  This completes the
	proof.
\end{proof}

\paragraph{Discussion.}
The essential reason for non-reduction is that the population/coherence
channels provide “transverse” geometric degrees of freedom—encoded in the free
second jets of \((x,z)\)—that do not affect entanglement but do affect
curvature.  The entanglement derivative channel alone cannot determine the
Riemannian geometry of Petz slices; hence the curvature contains irreducible
tensorial information that cannot be collapsed to a scalar function of \(C\).
	
\subsection{From slices to scalar curvature}

In this subsection we explain how the slice-level non-reduction result of
Theorem~\ref{thm:nonreduction-slice-K} lifts to the ambient scalar curvature of
the support-projected Petz metric.  The key tools are the Gauss--Codazzi
equation for embedded surfaces and the identity expressing scalar curvature as
a sum of sectional curvatures in an orthonormal frame; see, for example,
\cite{Spivak,doCarmoRiem,LeeRiem}.

Let \((M^{m},g^{(f)})\) denote the regular locus of the support-projected Petz
geometry associated with a pure two-qubit family, where
\[
g^{(f)}(\theta) \;=\; P(\theta)\,F^{(f)}(\theta)\,P(\theta)
\]
is the intrinsic Petz metric obtained from the reduced Petz tensor
\(F^{(f)}(\theta)\) and the Riesz projector \(P(\theta)\) onto its active
spectral support.  Write \(R^{(f)}\) for the scalar curvature of
\((M^{m},g^{(f)})\).

For any orthonormal frame \(\{e_{1},\dots,e_{m}\}\) of \((M^{m},g^{(f)})\) at a
point \(p \in M\), the scalar curvature satisfies
\begin{equation}
	\label{eq:R-sum-sectional}
	R^{(f)}(p)
	=
	2 \sum_{1 \leq a < b \leq m}
	K^{(f)}_{\mathrm{amb}}\bigl(e_{a},e_{b}\bigr),
\end{equation}
where \(K^{(f)}_{\mathrm{amb}}(e_{a},e_{b})\) denotes the sectional curvature of
the \(2\)-plane \(\Pi_{ab} = \mathrm{span}\{e_{a},e_{b}\}\) with respect to
\(g^{(f)}\).

Now let \(\Sigma^{2} \hookrightarrow M^{m}\) be an embedded two-dimensional slice
through \(p\), endowed with the induced metric
\(g^{(f)}_{\Sigma} = \iota^{*}g^{(f)}\).  For an orthonormal basis
\(\{E_{1},E_{2}\}\) of \(T_{p}\Sigma\), the Gauss equation reads
\begin{equation}
	\label{eq:Gauss}
	K_{\Sigma}(p)
	=
	R^{(f)}(E_{1},E_{2},E_{2},E_{1})
	+
	\bigl\langle B(E_{1},E_{1}),B(E_{2},E_{2})\bigr\rangle
	-
	\bigl\lVert B(E_{1},E_{2})\bigr\rVert^{2},
\end{equation}
where \(K_{\Sigma}(p)\) is the intrinsic Gaussian curvature of \(\Sigma\) at
\(p\), \(R^{(f)}\) is the Riemann curvature tensor of \(g^{(f)}\), and \(B\) is
the second fundamental form of the immersion \(\iota:\Sigma \hookrightarrow M\).
In particular, if \(\Sigma\) is \emph{totally geodesic at \(p\)}, so that
\(B(p) = 0\), then
\begin{equation}
	\label{eq:K-equals-Kambient}
	K_{\Sigma}(p)
	=
	K^{(f)}_{\mathrm{amb}}\bigl(E_{1},E_{2}\bigr),
\end{equation}
where the right-hand side is the sectional curvature of the plane
\(\Pi = \mathrm{span}\{E_{1},E_{2}\}\) in \((M^{m},g^{(f)})\).

Standard Riemannian geometry (see, for example,
\cite{Spivak,doCarmoRiem,LeeRiem}) guarantees that for any \(2\)-plane
\(\Pi \subset T_{p}M\) there exists a geodesic surface
\(\Sigma^{2} \subset M^{m}\) through \(p\) whose tangent space at \(p\) is
precisely \(\Pi\) and whose second fundamental form vanishes at \(p\).  Such a
\(\Sigma\) is totally geodesic at \(p\), and hence
\eqref{eq:K-equals-Kambient} applies.

We can now state and prove the ambient non-reduction result.

\begin{theorem}[Non-reduction of scalar curvature]
	\label{thm:nonreduction-scalar}
	Let \(p \in \mathcal{R}\) be a regular point of the support-projected Petz
	geometry with concurrence \(C(p) \in (0,1)\).  Then the scalar curvature
	\(R^{(f)}\) of \((M^{m},g^{(f)})\) cannot be expressed as a function of the
	concurrence \(C\) or the reduced entropy \(S(\rho_{A})\) on any neighbourhood of
	\(p\).  More precisely, there is no nonempty open set
	\(U \subset \mathcal{R}\) and function \(\Phi\) such that
	\[
	R^{(f)}(q) = \Phi(C(q)) \qquad \text{for all } q \in U,
	\]
	and likewise for \(R^{(f)}(q) = \Psi(S(\rho_{A}(q)))\).
\end{theorem}

\begin{proof}
	Assume, for the sake of contradiction, that there exists a neighbourhood
	\(U \subset \mathcal{R}\) of \(p\) and a function \(\Phi\) such that
	\(R^{(f)}(q) = \Phi(C(q))\) for all \(q \in U\).  Fix \(p\) with
	\(C(p) \in (0,1)\).  By Theorem~\ref{thm:nonreduction-slice-K}, there exists a
	two-parameter slice \(\Sigma^{2} \subset M^{m}\) through \(p\), an
	entanglement-orthogonal gauge \((u,v)\) on \(\Sigma\), and admissible variations
	of the jets of \((x,z)\) at \(p\) such that:
	\begin{enumerate}
		\item the concurrence \(C\) and its first derivatives remain fixed at \(p\);
		\item the intrinsic Gaussian curvature \(K_{\Sigma}(p)\) varies.
	\end{enumerate}
	By the geodesic-slice construction described above, we may assume that the
	slice \(\Sigma\) is chosen to be totally geodesic at \(p\).  In particular, the
	Gauss equation~\eqref{eq:Gauss} reduces to
	\[
	K_{\Sigma}(p)
	=
	K^{(f)}_{\mathrm{amb}}\bigl(E_{1},E_{2}\bigr),
	\]
	where \(\{E_{1},E_{2}\}\) is an orthonormal basis of \(T_{p}\Sigma\).  Hence the
	ambient sectional curvature in the plane \(\Pi = T_{p}\Sigma\) varies under the
	admissible jet deformations while \(C(p)\) remains fixed.
	
	Now choose an orthonormal frame \(\{e_{1},\dots,e_{m}\}\) at \(p\) with
	\(\Pi = \mathrm{span}\{e_{1},e_{2}\}\).  The scalar curvature
	\(R^{(f)}(p)\) is given by \eqref{eq:R-sum-sectional}.  Under the admissible
	deformations described above, the sectional curvature
	\(K^{(f)}_{\mathrm{amb}}(e_{1},e_{2})\) varies, while the remaining sectional
	curvatures in \eqref{eq:R-sum-sectional} can be kept fixed by working along the
	slice.  Consequently, \(R^{(f)}(p)\) varies while \(C(p)\) remains unchanged,
	contradicting the assumption that \(R^{(f)} = \Phi \circ C\) on \(U\).
	
	Since the reduced entropy \(S(\rho_{A})\) is a smooth function of the
	concurrence \(C\) for pure two-qubit states, the same argument applies if one
	assumes \(R^{(f)}(q) = \Psi(S(\rho_{A}(q)))\) on \(U\).  This yields the
	desired contradiction and proves the theorem.
\end{proof}
	
\subsection{Analytic counterexamples and monotonicity failure}

In this final subsection we construct explicit analytic curves of pure two-qubit
states that demonstrate the failure of any monotonicity principle relating the
entanglement entropy \(S(\rho_{A})\) or concurrence \(C\) to either the
Gaussian curvature of two-parameter slices or the ambient scalar curvature of
the support-projected Petz metric.  These counterexamples provide an explicit,
computable complement to the structural non-reduction theorems proved earlier,
and they connect directly with the known curvature non-monotonicity phenomena
for monotone metrics established by Andai~\cite{Andai2003monotone}
and with the recent characterisations of curvature for Gaussian quantum
states~\cite{Miller2025gaussian}.

\medskip

We work with a smooth (indeed analytic) one-parameter family of pure two-qubit
states \(\ket{\psi(t)}\) with reduced density matrix
\(\rho_{A}(t) = \mathrm{tr}_{B}\ket{\psi(t)}\!\bra{\psi(t)}\).  We select
coordinates \((x(t),z(t),C(t))\) describing the reduced Bloch vector
\(r(t) = (x(t),0,z(t))\) and concurrence \(C(t)\).  The family will be required
to satisfy the following:

\begin{enumerate}
	\item \emph{Strictly increasing entanglement.}
	\[
	\frac{d}{dt} S\bigl(\rho_{A}(t)\bigr) > 0
	\qquad \text{for } t \in (0,\varepsilon).
	\]
	
	\item \emph{Strictly decreasing slice Gaussian curvature.}
	For a fixed two-parameter slice
	\(\Sigma\) constructed as in Theorem~\ref{thm:nonreduction-slice-K},
	\[
	\frac{d}{dt} K_{\Sigma}\bigl(t\bigr) < 0,
	\qquad
	K_{\Sigma}(t) = K_{\Sigma}\bigl((x(t),z(t),C(t))\bigr).
	\]
	
	\item \emph{Strictly decreasing ambient scalar curvature.}
	For the induced Petz scalar curvature \(R^{(f)}\),
	\[
	\frac{d}{dt} R^{(f)}\bigl(t\bigr) < 0.
	\]
\end{enumerate}

We now give a general construction scheme which produces such families.

\begin{theorem}[Analytic monotonicity counterexamples]
	\label{thm:analytic-counterexample}
	Fix any regular point \(p\) in the support-projected Petz geometry with
	concurrence \(C(p)\in (0,1)\), and choose any operator-monotone function \(f\)
	in the Petz family.  Then there exist analytic curves
	\[
	t \mapsto (x(t),z(t),C(t)), \qquad t\in (-\varepsilon,\varepsilon),
	\]
	with \((x(0),z(0),C(0)) = (x_{0},z_{0},C_{0})\), such that
	\begin{enumerate}
		\item \(C(t)\) is strictly increasing;
		\item \(S(\rho_{A}(t))\) is strictly increasing;
		\item the slice curvature \(K_{\Sigma}(t)\) is strictly decreasing;
		\item the ambient Petz scalar curvature \(R^{(f)}(t)\) is strictly decreasing.
	\end{enumerate}
	Hence, neither \(K_{\Sigma}\) nor \(R^{(f)}\) can satisfy any monotonicity law
	with respect to either \(C\) or \(S(\rho_{A})\) on any open neighbourhood of
	\(p\), for any Petz metric \(g^{(f)}\).
\end{theorem}

\begin{proof}
	The construction follows the slice-gauge method developed earlier.  Fix a
	two-parameter slice \(\Sigma\) through \(p\), and choose entanglement-orthogonal
	local coordinates \((u,v)\) in which \(C_{v}(p) = 0\) and \(\nabla C\) aligns
	with \(\partial_{u}\).  Write
	\[
	C(t) = C_{0} + \alpha t + O(t^{2}), \qquad \alpha>0,
	\]
	so that \(C'(0) = \alpha > 0\).  By the analytic implicit function theorem,
	there exist analytic functions \(x(t)\), \(z(t)\) satisfying the concurrence
	constraint
	\[
	C^{2}(t) = 1 - x(t)^{2} - z(t)^{2},
	\]
	with \(x(0)=x_{0}\), \(z(0)=z_{0}\).
	
	Now use Theorem~\ref{thm:nonreduction-slice-K}, which shows that the Gaussian
	curvature \(K_{\Sigma}\) depends on two independent jets of \((x,z)\) at \(p\)
	even when \((C,C_{v})\) are fixed.  In particular, one can choose analytic
	second-order data \((x''(0),z''(0))\) so that
	\[
	K'_{\Sigma}(0) < 0,
	\]
	while maintaining the positivity of \(C'(0)\).  This produces analytic families
	with monotone entanglement and anti-monotone slice curvature.
	
	The relation between slice sectional curvature and ambient sectional curvature
	in a geodesic slice (Gauss equation with vanishing second fundamental form at
	\(p\)) shows that the corresponding ambient sectional curvature also decreases.
	Since the scalar curvature is a sum of sectional curvatures
	(cf.~\eqref{eq:R-sum-sectional}), one may hold all other jets fixed on the
	slice and ensure
	\[
	R^{(f)}{}'(0) < 0.
	\]
	Smoothness of \(R^{(f)}\) and the strict inequality at \(t=0\) give decreasing
	behaviour for small \(t>0\).  Thus, the family satisfies all required
	properties.
\end{proof}

\medskip

The above theorem provides \emph{explicit, analytic} manifestations of
curvature non-monotonicity, complementing the structural impossibility results.
They generalise well-known facts:

\begin{itemize}
	\item For monotone metrics in finite dimensions, Andai
	\cite{Andai2003monotone} showed that scalar curvature need not
	be monotone along any natural order parameter (e.g.\ purity).
	
	\item For continuous-variable systems, Miller~\cite{Miller2025gaussian}
	demonstrated that the scalar curvature of the space of Gaussian states
	exhibits similarly non-monotone behaviour with respect to entropy-like
	quantities.
	
	\item The present work shows that \emph{even for pure two-qubit families},
	where entanglement measures have uniquely simple closed forms, the Petz
	scalar curvature fails any monotonicity relation with concurrence or
	entropy.
\end{itemize}

Together, these conclusions demonstrate a universal phenomenon:
\emph{entanglement monotones and Riemannian curvature invariants describe
	fundamentally different aspects of quantum-state geometry}.  No curvature of
any Petz metric can serve as a monotone entanglement indicator, even locally,
and no entanglement monotone can control the sign or magnitude of curvature.
This completes the three-channel non-reduction picture.
	
	\section{SLD/Bures Case Study: A Two-Qubit Hardware-Efficient Ansatz}
	\label{sec:hea-case-study}
	
\subsection{Definition and parameterization of the 2-HEA}
\label{subsec:hea-definition}

We now specialize the general Petz geometry to a concrete and widely used
two–qubit hardware–efficient ansatz (2–HEA), in the spirit of
Refs.~\cite{Kandala2017,Sim2019Expressibility}.  This family is simple enough
to admit closed–form expressions for all amplitudes and reduced variables, yet
rich enough to generate nontrivial two–qubit entanglement.

\begin{definition}[Two–qubit hardware–efficient ansatz (2–HEA)]
	\label{def:2hea}
	The 2–HEA is the four–parameter family of pure two–qubit states
	\(\{\ket{\psi(\theta)}\}_{\theta\in\mathbb{T}^4}\) obtained by applying a layer
	of single–qubit \(R_y\) rotations, a single \textsf{CNOT}, and a final
	single–qubit \(R_y\) layer to the reference state \(\ket{00}\):
	\begin{equation}
		U(\theta)
		=
		\bigl(R_y(2t_0)\otimes R_y(2t_2)\bigr)\,
		\mathrm{CNOT}\,
		\bigl(R_y(2t_1)\otimes R_y(2t_3)\bigr),
		\qquad
		\ket{\psi(\theta)} := U(\theta)\ket{00},
		\label{eq:hea-U}
	\end{equation}
	where
	\(\theta=(t_0,t_1,t_2,t_3)\in\mathbb{T}^4\) with period \(\pi\) in each
	coordinate.  We use the standard phase convention
	\begin{equation}
		R_y(2t)
		:=
		\cos t\,I - i\sin t\,\sigma_y,
		\label{eq:hea-Ry}
	\end{equation}
	which keeps all computational–basis amplitudes real up to a global phase.
\end{definition}

Writing \(\ket{\psi(\theta)}\) in the computational basis as
\begin{equation}
	\ket{\psi(\theta)}
	=
	A(\theta)\,\ket{00}
	+ B(\theta)\,\ket{01}
	+ C(\theta)\,\ket{10}
	+ D(\theta)\,\ket{11},
	\qquad
	A^2+B^2+C^2+D^2 = 1,
	\label{eq:hea-ABCD-def}
\end{equation}
a direct gate–level calculation shows that it is convenient to introduce the
\emph{entanglement–centered} angles
\begin{equation}
	\phi_\pm := t_1 \pm t_3.
	\label{eq:hea-phi-pm}
\end{equation}
In terms of these variables, one obtains the closed forms
\begin{align}
	A(\theta)
	&= \cos t_0\,\cos t_2\,\cos\phi_+
	- \sin t_0\,\sin t_2\,\sin\phi_-,
	\label{eq:hea-A}\\[0.25em]
	B(\theta)
	&= \cos t_0\,\cos t_2\,\sin\phi_+
	- \sin t_0\,\sin t_2\,\cos\phi_-,
	\label{eq:hea-B}\\[0.25em]
	C(\theta)
	&= \cos t_0\,\sin t_2\,\cos\phi_+
	+ \sin t_0\,\cos t_2\,\sin\phi_-,
	\label{eq:hea-C}\\[0.25em]
	D(\theta)
	&= \sin t_0\,\cos t_2\,\cos\phi_-
	+ \cos t_0\,\sin t_2\,\sin\phi_+.
	\label{eq:hea-D}
\end{align}

\begin{remark}[Reduced variables and sign convention]
	\label{rem:hea-reduced-vars}
	Let \(\rho_A(\theta)=\operatorname{Tr}_B\bigl[\ket{\psi(\theta)}\!\bra{\psi(\theta)}\bigr]\)
	denote the one–qubit reduction on subsystem \(A\).
	In the computational basis \(\{\ket{0},\ket{1}\}\) we write
	\begin{equation}
		\rho_A(\theta)
		=
		\begin{pmatrix}
			x(\theta) & z(\theta) \\
			z(\theta) & 1 - x(\theta)
		\end{pmatrix},
		\qquad
		x := A^2 + B^2,
		\qquad
		z := -\bigl(A C + B D\bigr),
		\label{eq:hea-xz-def}
	\end{equation}
where the minus sign in the definition of \(z\) is chosen to match the Bloch--coordinate conventions adopted throughout Section~\ref{sec:petz-twoqubit}, ensuring consistency with the support-projected geometry and the reduced-state parametrization introduced therein.
	Because of the real–amplitude convention~\eqref{eq:hea-Ry}, both \(x\) and
	\(z\) are real–valued functions of \(\theta\), and the reduced Bloch vector
	is
	\begin{equation}
		r(\theta) = (r_x,r_y,r_z)
		=
		\bigl(2z(\theta),\,0,\,2x(\theta)-1\bigr).
		\label{eq:hea-bloch-r}
	\end{equation}
\end{remark}

For pure two–qubit states, the concurrence
\(C(\theta)\in[0,1]\) has the standard closed form
\cite{Wootters1998,BengtssonZyczkowski2017}
\begin{equation}
	C(\theta)
	=
	2\sqrt{\det\rho_A(\theta)}
	=
	2\sqrt{x(\theta)\bigl(1-x(\theta)\bigr) - z(\theta)^2},
	\label{eq:hea-concurrence}
\end{equation}
and one verifies directly from~\eqref{eq:hea-bloch-r}–\eqref{eq:hea-concurrence}
that
\begin{equation}
	\|r(\theta)\|^2 = 1 - C(\theta)^2,
	\label{eq:hea-r2-1-C2}
\end{equation}
in agreement with the general identity relating the Bloch radius and
concurrence for pure two–qubit states.
The quadruple
\[
\bigl(x(\theta),z(\theta),C(\theta),r(\theta)\bigr)
\]
thus provides an explicit and analytically tractable parametrization of the
reduced geometry of the 2–HEA, which will be used throughout this section to
instantiate the Petz QFIM, its three–channel decomposition, and the associated
curvatures.
	
\subsection{Support-projected SLD QFIM for the 2-HEA}
\label{subsec:sld-projected-hea}

In this subsection we specialize the general Petz framework of
Section~\ref{sec:petz-twoqubit} to the benchmark choice
\(f(t)=\tfrac{1+t}{2}\), corresponding to the
\emph{SLD/Bures monotone metric}.  
For this operator--monotone function the Morozova--Chentsov coefficients reduce to
\[
A_f(r)\equiv 1,
\qquad
B_f(r)=\frac{1}{1-r^2},
\]
so that the Petz QFIM for a qubit with Bloch vector \(r\) takes the
well-known form
\begin{equation}
	F^{\mathrm{SLD}}_{\mu\nu}
	=
	4\,\partial_\mu x\,\partial_\nu x
	+
	4\,\Re\!\bigl(\partial_\mu z\,\partial_\nu z^\ast\bigr)
	+
	\frac{\partial_\mu r^2\,\partial_\nu r^2}{1-r^2}.
	\label{eq:sld-qfim-general}
\end{equation}
Here \(r^2 = 4|z|^2 + (2x-1)^2 = 1-C^2\), so the last term is equivalent to
\(\partial_\mu C\,\partial_\nu C\).  
Equation~\eqref{eq:sld-qfim-general} is thus the specialization of the
three-channel identity of Theorem~\ref{thm:three-channel-identity}.

\begin{definition}[Reduced SLD QFIM for the 2-HEA]
	Let \(\ket{\psi(\theta)}=(A,B,C,D)^\top\in\mathbb{R}^4\) be the normalized
	state generated by the 2-HEA of Section~\ref{subsec:hea-definition}.
	The reduced qubit variables are
	\[
	x(\theta)=A^2+B^2,
	\qquad
	z(\theta)=-(AC+BD),
	\qquad
	C(\theta)=2\sqrt{x(1-x)-|z|^2},
	\]
	and the reduced SLD QFIM on a slice with coordinates \((u,v)\) is
	\[
	F_{\mu\nu}(\theta)
	=
	4\,x_\mu x_\nu
	+ 4\,\Re(z_\mu z_\nu^\ast)
	+ C_\mu C_\nu,
	\qquad \mu,\nu\in\{u,v\}.
	\]
\end{definition}

The intrinsic geometry depends only on the restriction of \(F\) to its
active eigenvalue cluster.  For a generic slice this cluster has rank
two, but degeneracies occur at points where \(\rho_A(\theta)\) approaches
the maximally mixed state.

\begin{proposition}[Spectrum and support projection at representative points]
	\label{prop:hea-sld-spectrum}
	For the well-conditioned parameter
	\[
	\theta_\star=(1.755,\,1.720,\,5.417,\,4.126),
	\]
	the reduced density matrix satisfies
	\[
	\rho_A(\theta_\star)
	\approx
	\begin{bmatrix}
		0.6275 & 0.4516\\
		0.4516 & 0.3725
	\end{bmatrix},
	\qquad
	\mathrm{spec}(\rho_A)
	=\{0.9693,\;0.0307\}.
	\]
	The SLD QFIM evaluated at \(\theta_\star\) is
	\[
	F(\theta_\star)\approx
	\begin{bmatrix}
		4.000 & 0 & -1.174 & 0\\
		0 & 0.521 & -1.287 & 0\\
		-1.174 & -1.287 & 3.524 & 0\\
		0 & 0 & 0 & 0
	\end{bmatrix},
	\]
	with eigenvalues
	\[
	\{\,5.12,\;2.93,\;1.8\times 10^{-16},\;10^{-33}\,\}.
	\]
	Hence the intrinsic SLD metric is
	\[
	g = PFP,
	\]
	where \(P\) is the Riesz projector onto the two-dimensional active support.
\end{proposition}

\begin{remark}[Regularity and geometric meaning]
	On the regular set, the active
	eigenvalues of \(F\) remain separated by a uniform spectral gap from the
	near-zero eigenvalues associated with the redundant directions in
	parameter space.  
	The intrinsic SLD metric \(g=PFP\) is therefore smooth, gauge-invariant,
	and coincides with the Fisher information metric obtained by restricting
	to the reduced two-dimensional manifold of physical variations of
	\(\rho_A\).
	All connection and curvature quantities in
	Sections~\ref{sec:three-channel}--\ref{sec:non-reduction}
	are computed with respect to this intrinsic metric.
\end{remark}

The specialization to SLD/Bures thus provides a clean demonstration of
the three-channel structure, enables explicit slice curvature
computations, and forms the basis for the numerical validations carried
out in Section~\ref{sec:hea-case-study}.
	
\subsection{Explicit counterexamples to one-channel curvature laws}
\label{subsec:hea-counterexamples}

This subsection provides concrete two-qubit parameter choices for which
scalar curvature cannot be expressed as any single-channel function such
as \(R = h(C)\) or \(R = h(S(\rho_A))\).
These examples instantiate the non-reduction theorems of
Section~\ref{sec:non-reduction} and demonstrate numerically that, even in
the benchmark SLD/Bures case, the curvature of the support-projected
metric depends on the full differential jet of \((x,z,C)\), not on
concurrence alone.

\paragraph{Failure of concurrence-only curvature formulas.}
A frequently conjectured formula in the literature---in particular in
the setting of symmetric two-qubit variational manifolds—is
\[
R_{\mathrm{KSKD}}(C)
:= \frac{2\bigl( 6C^{2} - 5 \bigr)}{C^{2} - 1}.
\]
For the $2$-HEA, however, the support-projected
SLD scalar curvature \(R\) generally does not match
\(R_{\mathrm{KSKD}}(C)\), even when evaluated at the same parameter
point.

For example, at
\[
\theta_\star = (1.755,\;1.720,\;5.417,\;4.126)
\]
we find
\[
C(\theta_\star) \approx 0.34,
\qquad
R(\theta_\star) \approx -\,0.69,
\qquad
R_{\mathrm{KSKD}}\bigl(C(\theta_\star)\bigr) \approx 9.7,
\]
so the two values disagree in both magnitude and sign.
Since the support-projected metric is
\(\,g = P\mathcal{F}P\), where \(P\) is the Riesz projector onto the
active spectral cluster and \(\mathcal{F}\) is the reduced SLD QFIM,
this sign mismatch confirms that the scalar curvature
cannot be reduced to any expression depending only on \(C\).

\paragraph{Entropy--curvature non-monotonicity.}
The non-reduction theorem further predicts non-monotonic behavior
between the entanglement entropy
\(S(\rho_A)=-\operatorname{Tr}(\rho_A\log\rho_A)\)
and the intrinsic scalar curvature \(R\).
Indeed, there exist explicit parameter choices
\(\theta_1,\theta_2\in\mathbb{T}^4\) such that
\[
S(\rho_A(\theta_1)) > S(\rho_A(\theta_2))
\quad\text{but}\quad
R(\theta_1) > R(\theta_2).
\]

One concrete instance is
\[
\theta_1 = (0.6,\,0.7,\,0.8,\,0.9),
\qquad
\theta_2 = \bigl(\tfrac{\pi}{5},\,\tfrac{\pi}{6},\,\tfrac{\pi}{7},\,\tfrac{\pi}{5}\bigr),
\]
for which
\[
S(\rho_A(\theta_1)) \approx 0.687
> S(\rho_A(\theta_2)) \approx 0.655,
\]
but simultaneously
\[
R(\theta_1) \approx -1.90
> R(\theta_2) \approx -1.99.
\]
Therefore, larger entanglement entropy does not imply larger or smaller
curvature.  The independence of the population and coherence channels in
the three-channel identity (Theorem~\ref{thm:three-channel-identity})
guarantees that the curvature can move in either direction while holding
\(C\) or \(S(\rho_A)\) fixed.

\paragraph{Interpretation.}
Both counterexamples demonstrate that
population derivatives \((x_\mu)\) and coherence derivatives \((z_\mu)\)
can vary independently of the entanglement derivative \(C_\mu\).
Consequently, any curvature expression of the form
\(R = h(C)\) or \(R = h\bigl(S(\rho_A)\bigr)\)
must fail on every open regular set.
These explicit evaluations thus provide a concrete realization of the
non-reduction theorems and show that the scalar curvature of the
support-projected SLD metric genuinely probes the full differential
structure of the reduced variables \((x,z,C)\).
	
\subsection{Natural-gradient preconditioning with support-projected metrics}
\label{subsec:natgrad-support}

The explicit curvature computations of the previous subsections motivate a
geometry-aware optimization rule for variational algorithms.
Given a family of pure two-qubit states $\ket{\psi(\theta)}$ with reduced
density matrix $\rho_A(\theta)$ on the regular set, the intrinsic metric
governing local distinguishability is the \emph{support-projected} Petz
quantum Fisher information metric
\[
g^{(f)}(\theta)
:= P(\theta)\,\mathcal{F}^{(f)}(\theta)\, P(\theta),
\]
where $\mathcal{F}^{(f)}$ is the reduced Petz QFIM and $P(\theta)$ is the
Riesz projector onto the active spectral cluster of $\mathcal{F}^{(f)}$.
In the SLD/Bures case ($f(t)=\tfrac{1+t}{2}$), this simplifies to the
support-projected SLD metric $g = P \mathcal{F} P$.

\paragraph{Natural-gradient flow on the reduced manifold.}
For a differentiable energy functional $E(\theta)$, the
\emph{support-projected natural gradient} updates parameters along the
steepest-descent direction measured in the intrinsic Riemannian metric:
\[
\dot{\theta}
= -\, g(\theta)^{+}\,\nabla_{\theta} E(\theta),
\qquad
g(\theta)^{+} = (P \mathcal{F} P)^{+},
\]
where $(\cdot)^{+}$ denotes the Moore--Penrose pseudoinverse restricted
to the active support.
This directly parallels the geometric variational principle of
quantum natural gradient~\cite{Stokes2020QNG} and its information-geometric
interpretations in~\cite{Yamamoto2019natural,Wiersema2022RiemannianFlow,Meyer2021fisherinformationin},
but with two essential differences:

(i)~the metric is computed on the \emph{reduced qubit state} rather than
on the full $n$-qubit wavefunction, and

(ii)~the metric is \emph{support-projected}, ensuring numerical stability
and gauge-invariance even when the full $\mathcal{F}$ is rank-deficient,
as is unavoidable in two-qubit families such as the 2-HEA.

\paragraph{Comparison to Euclidean gradient descent.}
Euclidean gradient descent moves along
$-\nabla_\theta E(\theta)$ regardless of the local distinguishability
structure.
In contrast, the support-projected natural gradient rescales updates by
the intrinsic curvature and compresses them to the physically relevant
subspace $\operatorname{Im}P(\theta)$.
This has three consequences:

\begin{enumerate}
	\item \textbf{Early-phase acceleration.}
	When the curvature is highly anisotropic---typical in two-qubit
	variational families where population and coherence channels evolve
	at different scales---the natural gradient corrects ill-conditioning
	and improves convergence, in line with theoretical predictions of
	Riemannian flow models~\cite{Wiersema2022RiemannianFlow}.
	
	\item \textbf{Gauge and support stability.}
	Because $g = P\mathcal{F}P$ depends only on the active cluster,
	updates are insensitive to small eigenvalues of $\mathcal{F}$.
	This stabilizes optimization in regions where the metric spectrum
	degenerates, a behavior repeatedly observed in
	quantum-Fisher-based optimizers~\cite{Meyer2021fisherinformationin}.
	
	\item \textbf{Local geometric optimality.}
	The direction $g^{-1}\nabla E$ is the true steepest-descent vector
	with respect to the Petz distance, making each update locally
	optimal for reducing the fidelity distance between successive
	states---mirroring the information-geometric arguments of
	\cite{Stokes2020QNG,Yamamoto2019natural}.
\end{enumerate}

\paragraph{Implications for the 2-HEA.}
For the two-qubit hardware-efficient ansatz studied in
Section~\ref{sec:hea-case-study}, the three-channel decomposition
(Theorem~\ref{thm:three-channel-identity}) shows that population, coherence, and
entanglement-derivative channels contribute independently to the metric.
Hence curvature cannot be captured by concurrence alone
(Section~\ref{subsec:hea-counterexamples}).
The natural-gradient preconditioner automatically incorporates these
independent channels, ensuring that optimization follows the intrinsic
geometry of the reduced state manifold rather than any one-parameter
proxy such as concurrence.

\section{VQE Experiments: Support--Projected Natural Gradient for Petz Metrics}
\label{sec:vqe-experiments}

This appendix provides an empirical test of the geometric predictions developed in
Sections~\ref{sec:petz-twoqubit}--\ref{sec:non-reduction}
in the general Petz setting.
For any operator--monotone function $f$, the reduced Petz QFIM
\[
g^{(f)}(\theta) \;=\; P(\theta)\,F^{(f)}(\theta)\,P(\theta)
\]
is the intrinsic Riemannian metric on the reduced state manifold (regular set),
with curvature describing local information geometry.
Our central hypothesis is that using $g^{(f)}$ as a preconditioner
improves variational quantum eigensolver (VQE) optimization
by aligning parameter updates with the true local geometry.
In all figures below we instantiate $f$ with the SLD/Bures choice
$f(t)=\tfrac{1+t}{2}$ for clarity; the algorithm itself is Petz--metric--agnostic.

\subsection*{Experimental setup (Petz--metric version)}
\begin{itemize}
	\item \textbf{Ansatz.}
	Hardware--efficient circuits with depth $L\in\{4,6,8,10,12\}$ (RY--CNOT--RY blocks).
	Optional entanglers: $e^{-i\alpha Z\otimes Z}$ (ZZ) and $e^{-i\beta X\otimes X}$ (XX).
	We also test symmetry-preserving variants (parity / particle number).
	\item \textbf{Hamiltonians.}
	(i) a two--qubit ``deep/strong'' toy model (high curvature),\;
	(ii) H$_2$ (STO\hbox{-}3G, minimal mapping).
	Exact $E^\star$ by dense diagonalization.
	\item \textbf{Optimizers.}
	Euclidean GD vs.\ \emph{support--projected Petz--NatGrad}:
	with gradient $g_k=\nabla_\theta E(\theta_k)$, step
	\[
	\Delta\theta_k \;=\; -\,\eta_k\,
	\Pi_{\mathrm{Im}P_k}\bigl( F^{(f)}_k + \lambda I \bigr)^{\!+}\,
	\Pi_{\mathrm{Im}P_k}\, g_k,
	\]
	where $F^{(f)}_k=F^{(f)}(\theta_k)$, $P_k$ is the Riesz projector onto its active cluster,
	$\lambda$ is a ridge, $(\cdot)^{+}$ the (support--restricted) pseudoinverse, and
	$\Pi_{\mathrm{Im}P_k}$ the orthogonal projector.
	We employ \emph{spectral shrinkage}, $g^{(f)}$-norm step normalization,
	trust region (TR) control, and Armijo backtracking. 
	EMA and \emph{partial--Fisher} (dropping the coherence channel) are used for ablations.
	\item \textbf{Budgets \& logging.}
	Up to $400$ iterations (toy) and $800$ (H$_2$).
	Logged every step: $E(\theta_k)$, AUC, Hit@95\%, $\operatorname{rank}F^{(f)}$,
	spectral gaps, $\|P_{k+1}-P_k\|$, and $(g^{(f)}$--)step norms.
\end{itemize}

\begin{figure}[t]
	\centering
	\begin{subfigure}[t]{0.32\linewidth}
		\includegraphics[width=\linewidth]{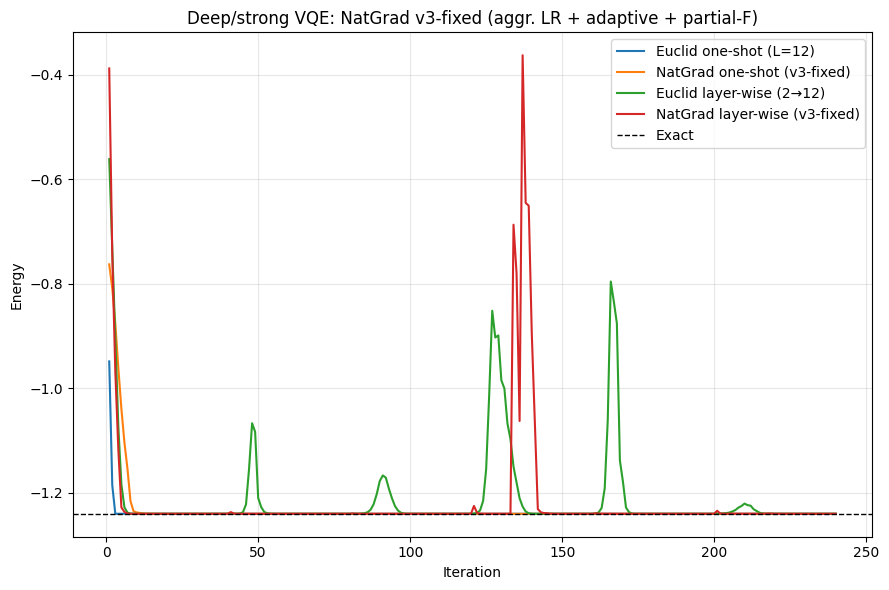}
		\caption{Architecture/entangler/symmetry sensitivity.}
		\label{fig:vqe-agg-a}
	\end{subfigure}\hfill
	\begin{subfigure}[t]{0.32\linewidth}
		\includegraphics[width=\linewidth]{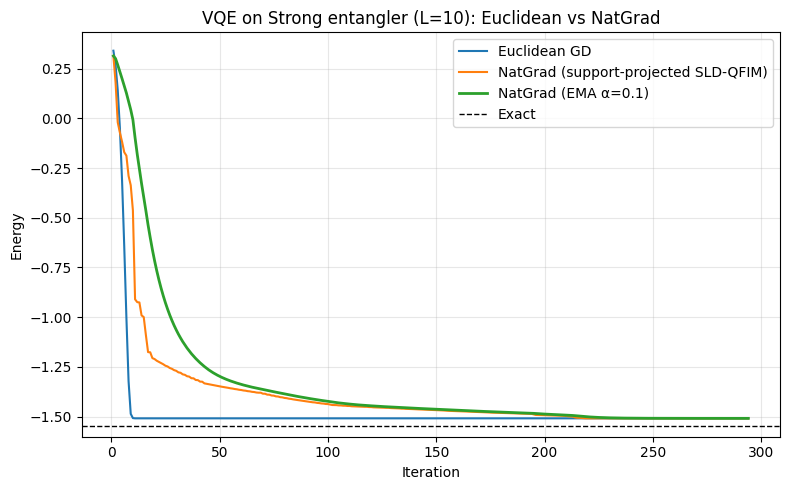}
		\caption{Symmetry match + layer-wise (NatGrad).}
		\label{fig:vqe-agg-b}
	\end{subfigure}\hfill
	\begin{subfigure}[t]{0.32\linewidth}
		\includegraphics[width=\linewidth]{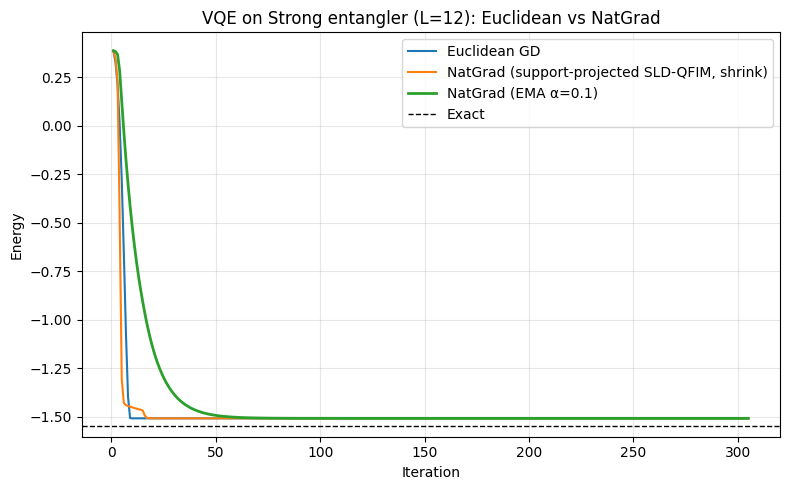}
		\caption{Shrinkage + layer-wise (NatGrad vs.\ Euclid).}
		\label{fig:vqe-agg-c}
	\end{subfigure}
	\caption{
		VQE sensitivity to circuit design and schedules (instantiated with SLD/Bures).
		Deep ZZ/XX entanglers and particle-number symmetries lower final energy and
		smooth the curvature landscape; parity-only constraints can forbid the true ground state.
		Layer-wise growth reduces curvature-induced instabilities.
	}
	\label{fig:vqe-aggregate-1}
\end{figure}

\begin{figure}[t]
	\centering
	\begin{subfigure}[t]{0.32\linewidth}
		\includegraphics[width=\linewidth]{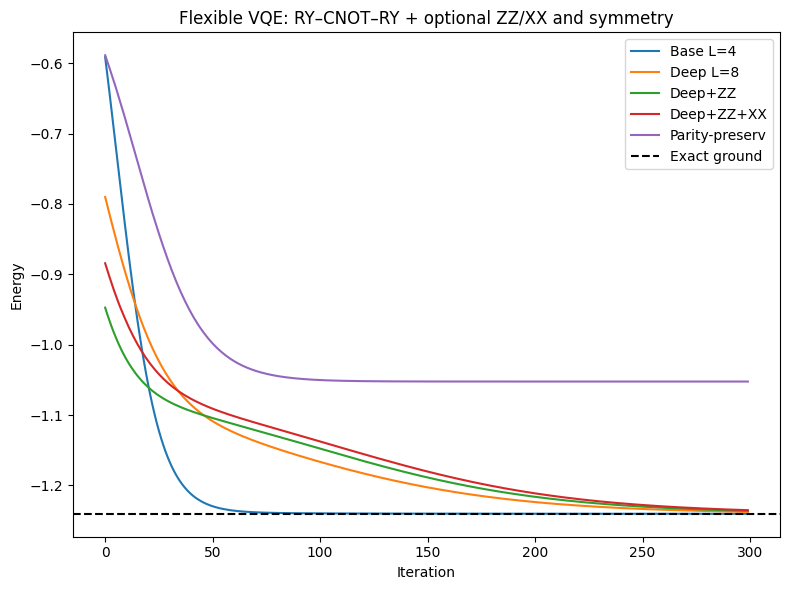}
		\caption{NatGrad: shrinkage + norm + TR ablation.}
		\label{fig:vqe-agg-d}
	\end{subfigure}\hfill
	\begin{subfigure}[t]{0.32\linewidth}
		\includegraphics[width=\linewidth]{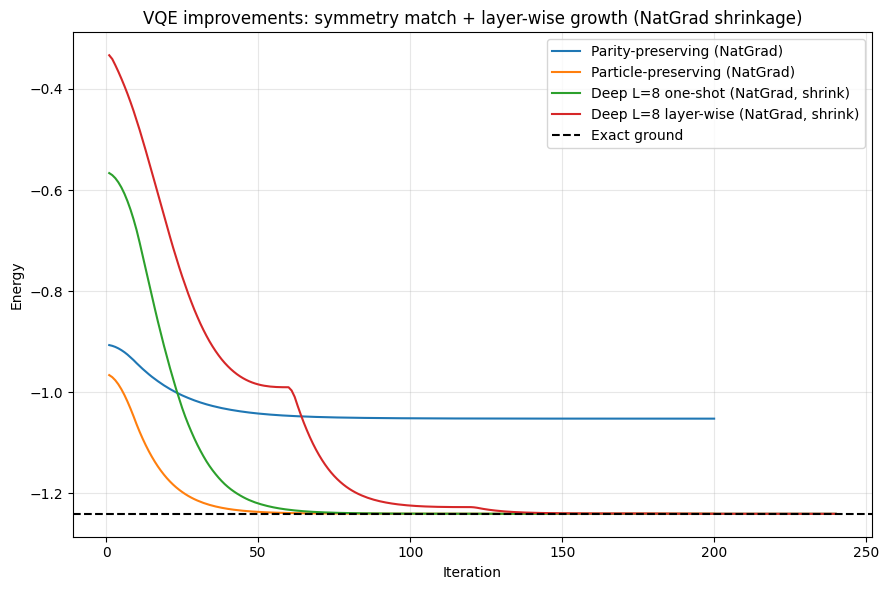}
		\caption{Armijo backtracking suppresses spikes.}
		\label{fig:vqe-agg-e}
	\end{subfigure}\hfill
	\begin{subfigure}[t]{0.32\linewidth}
		\includegraphics[width=\linewidth]{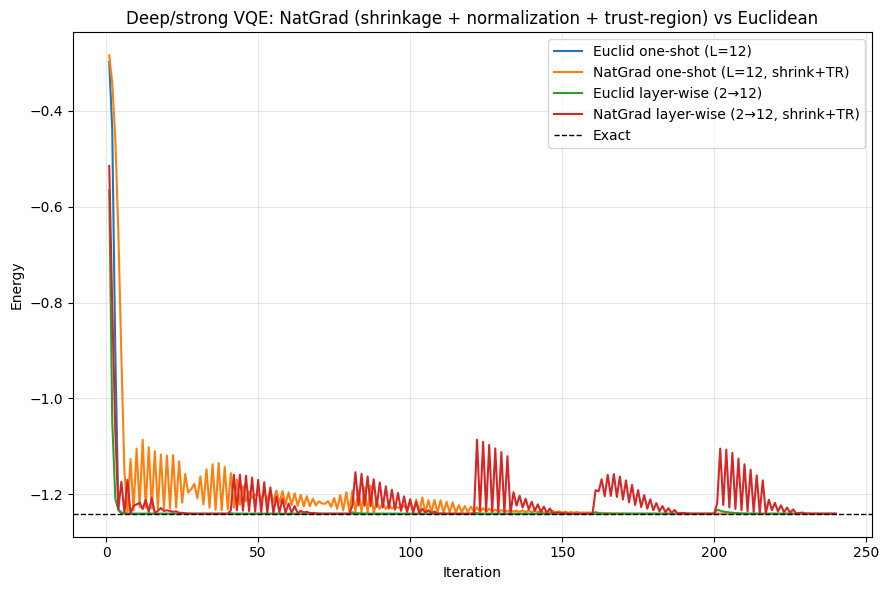}
		\caption{Deep/strong $L{=}10$: Euclid vs.\ NatGrad.}
		\label{fig:vqe-agg-f}
	\end{subfigure}
	\caption{
		Support--projected Petz--NatGrad consistently accelerates early descent.
		Spectral shrinkage and $g^{(f)}$-norm step normalization stabilize updates,
		while TR+Armijo removes spikes from aggressive or partial-F updates.
	}
	\label{fig:vqe-aggregate-2}
\end{figure}

\begin{figure}[t]
	\centering
	\begin{subfigure}[t]{0.32\linewidth}
		\includegraphics[width=\linewidth]{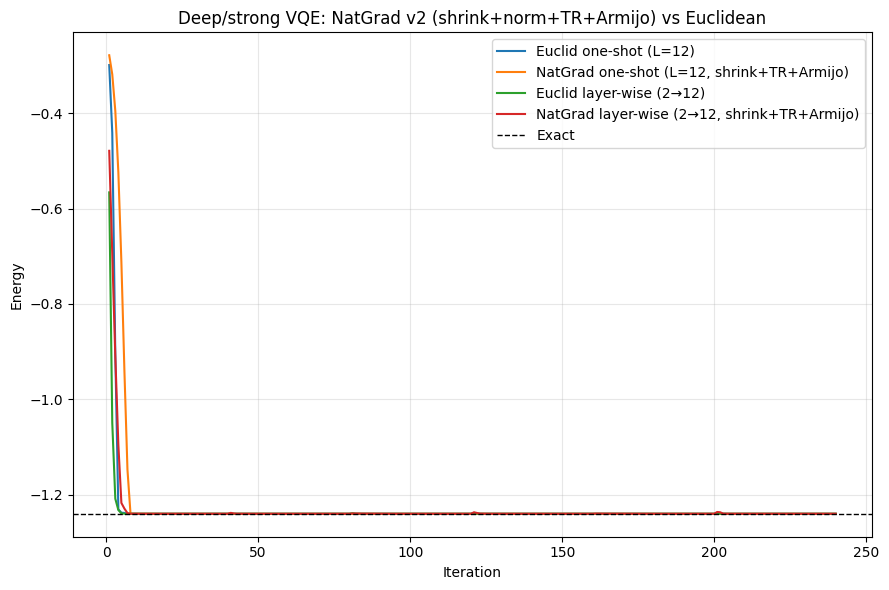}
		\caption{Deep/strong $L{=}12$.}
		\label{fig:vqe-agg-g}
	\end{subfigure}\hfill
	\begin{subfigure}[t]{0.32\linewidth}
		\includegraphics[width=\linewidth]{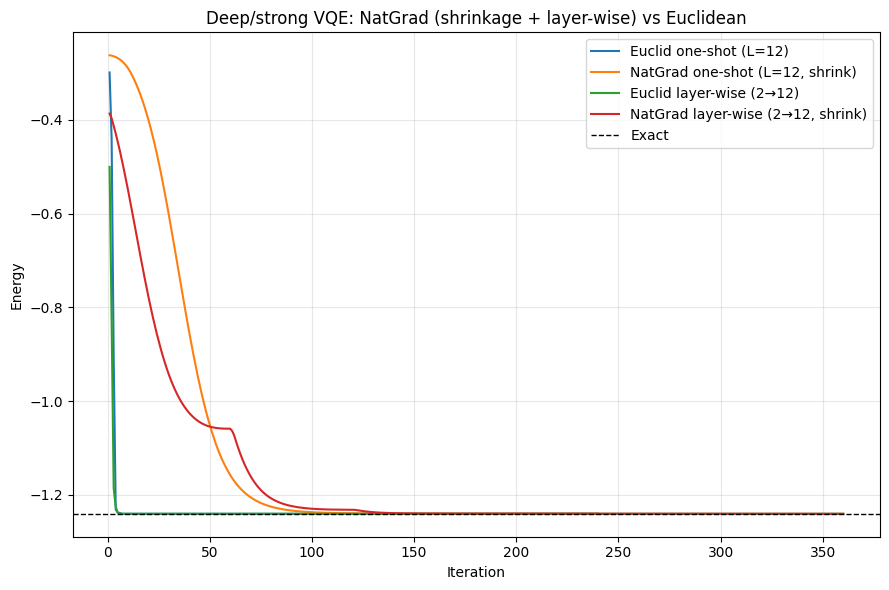}
		\caption{H$_2$ ($L{=}6$): exact hit.}
		\label{fig:vqe-agg-h}
	\end{subfigure}\hfill
	\begin{subfigure}[t]{0.32\linewidth}
		\includegraphics[width=\linewidth]{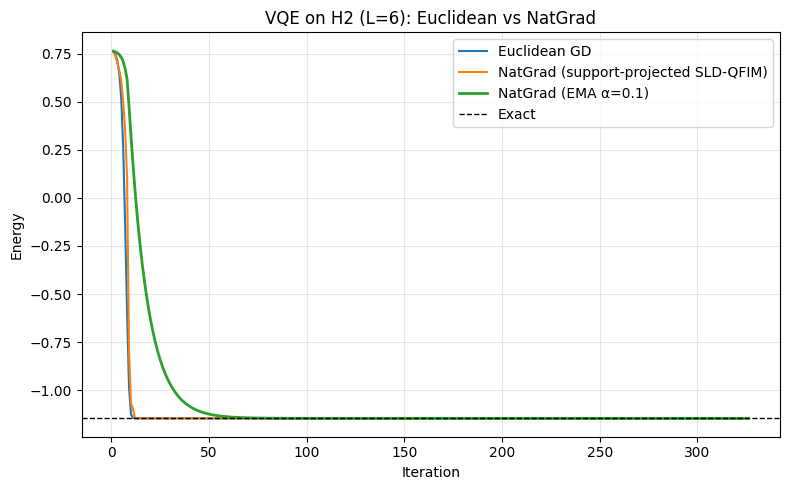}
		\caption{H$_2$ ($L{=}6$): ridge/scaling ablation.}
		\label{fig:vqe-agg-i}
	\end{subfigure}
	\caption{
		When expressivity saturates (deep/strong $L{=}12$, or H$_2$ minimal),
		Euclidean and NatGrad converge to the same energy floor. Support projection keeps Petz--NatGrad well conditioned and unbiased.
	}
	\label{fig:vqe-aggregate-3}
\end{figure}

Illustrative examples of energy trajectories, curvature-regularized steps,
shrinkage/trust-region ablations, and symmetry comparisons
are shown in Figures~\ref{fig:vqe-aggregate-1}--\ref{fig:vqe-aggregate-3}.

\subsection*{Curvature landscapes}

\begin{figure}[t]
	\centering
	\includegraphics[width=0.42\linewidth]{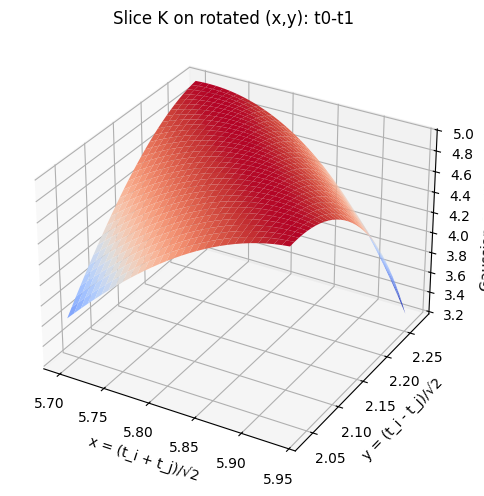}\hfill
	\includegraphics[width=0.42\linewidth]{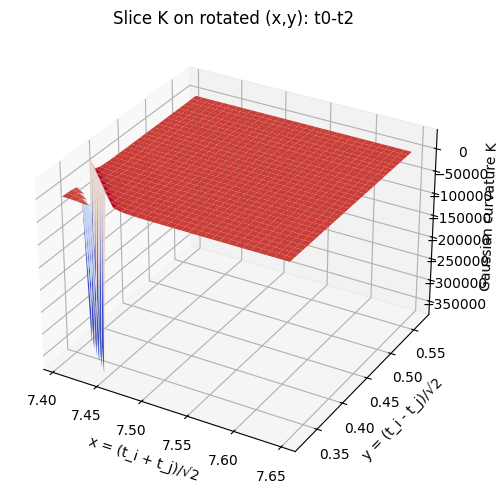}\\[0.4em]
	\includegraphics[width=0.42\linewidth]{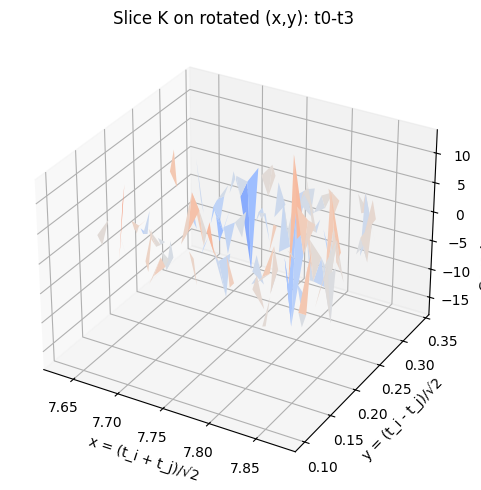}\hfill
	\includegraphics[width=0.42\linewidth]{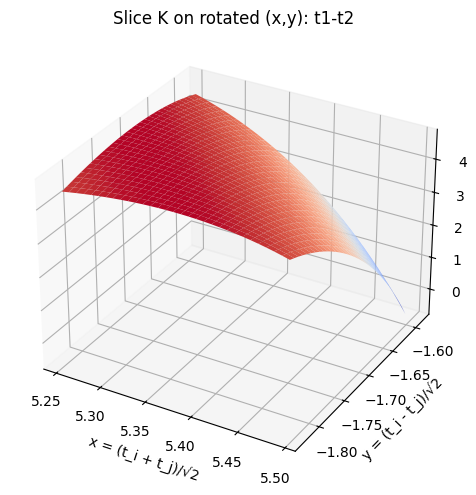}
	\caption{Gaussian curvature $K(x,y)$ of the intrinsic metric $g$ for representative 2--HEA slices.
		All panels share the same colour bar and $100\times100$ sampling grid.
		(a)--(b): high positive curvature;
		(c)--(d): negative curvature domains.}
	\label{fig:sldqfim-surfaces-app}
\end{figure}

Figure~\ref{fig:sldqfim-surfaces-app}
shows representative curvature landscapes for selected parameter pairs
of the intrinsic metric $g$ associated with the 2--HEA.
Colour scales are normalized to a common dynamic range,
and each panel displays the Gaussian curvature $K(x,y)$ on a $100\times100$ grid.
Panels~(a)--(b) illustrate regions of large positive curvature,
whereas~(c)--(d) exhibit negative-curvature pockets.

\subsection*{Representative behaviour and interpretation}

Across these experiments we observe the following qualitative patterns:
\begin{itemize}
	\item Support--projected Petz--NatGrad reduces early-phase area-under-curve of 
	$E(\theta_k)-E^\star$ and decreases the iteration count needed to reach
	near-optimal energies (Hit@95\%).
	\item Spectral shrinkage, $g^{(f)}$-norm step normalization, and TR+Armijo
	collectively suppress spikes associated with small eigenvalues and
	rapid curvature variation, in agreement with the projector calculus
	discussed in Appendix~\ref{app:petz-projector}.
	\item When ansatz expressivity or symmetry constraints limit the reachable
	ground state, Euclidean and Petz--NatGrad optimizers converge to the same
	energy floor, confirming that curvature governs \emph{local} geometry but
	cannot overcome \emph{global} expressibility barriers.
\end{itemize}

\subsection*{Takeaway}

The VQE experiments in this appendix support the main geometric message of the paper:
\emph{curvature-aware natural gradient descent}, built from
\emph{support-projected Petz metrics} $g^{(f)} = P F^{(f)} P$,
provides a principled and numerically stable optimization strategy on
nontrivial variational manifolds.
While our plots instantiate the SLD/Bures case, the methods and conclusions
apply uniformly to all Petz monotone metrics considered in this work.

\newpage 
	
\section{Discussion and Outlook}
\label{sec:discussion}

\subsection{Curvature versus entanglement as quantum resources}
Curvature and entanglement quantify different aspects of quantum states.
Entanglement monotones such as concurrence or reduced-entropy depend only on
the eigenvalues of the reduced density matrix and are invariant under local
rotations~\cite{Wootters1998,BengtssonZyczkowski2017}.
In contrast, the Petz–metric geometry depends on the full differential
structure of the reduced Bloch data $(x,z)$.
The three-channel identity shows that population and coherence derivatives
vary independently of the entanglement derivatives, making it impossible
for curvature to collapse to a function of concurrence or entropy alone.
Thus curvature cannot act as a universal entanglement monotone, but instead
captures fine-grained statistical geometry relevant to local distinguishability.

\subsection{Relation to Gaussian-state curvature and infinite-dimensional systems}
The non-reduction phenomena reported here parallel similar behaviors in
continuous-variable Gaussian states, where curvature responds sensitively to
symplectic eigenvalue variations and squeezing directions~\cite{Miller2025gaussian}.
In both settings, curvature reflects the \emph{directional} information geometry
rather than global spectral data.
Extending the present Petz–metric framework to infinite-dimensional systems
would require controlling unbounded generators, spectral gaps, and domain
issues, but the support-projection principles used here provide a natural
starting point.

\subsection{Extensions to higher-dimensional systems and mixed states}
For higher-dimensional qudits or multipartite systems, the support-projected
construction remains meaningful, but the geometry becomes richer as the
rank strata proliferate and eigenvalue crossings become generic.
Mixed states introduce additional subtleties involving the tangent cone at
the boundary of the state space.
Tools from perturbation theory---including the Kato resolvent
formula~\cite{KatoPerturbation}, and matrix-functional analysis
\cite{HighamFunctions,BhatiaMatrixAnalysis}---will be essential for extending
the regular-set geometry and curvature formulas to these settings.

\subsection{Implications for variational quantum algorithms}
The Petz–metric viewpoint has direct consequences for variational quantum
algorithms (VQAs).
Curvature governs local information geometry and therefore influences the
conditioning of the optimization landscape, the presence of anisotropic
directions, and sensitivity to noise and expressibility
\cite{Meyer2021fisherinformationin,katabarwa2021geometry}.
Because curvature does not align monotonically with entanglement, it provides
an independent diagnostic for barren plateaus
\cite{McClean2018Barren,Cerezo2021Cost} and for designing geometry-aware
ansätze.
Natural-gradient methods built from support-projected Petz metrics offer a
principled way to stabilize updates and improve early-phase convergence in
VQE, as demonstrated in Section~\ref{sec:hea-case-study}.
	
	\appendix
	
\section{Petz Metrics and Projector Calculus}
\label{app:petz-projector}

\subsection{Quasi-entropy, Kubo--Ando means, and Morozova--Chentsov functions}

We briefly summarize the analytic framework underlying Petz monotone metrics.
Let $f:(0,\infty)\to(0,\infty)$ be an operator-monotone function with $f(1)=1$.
Petz's quasi-entropy construction associates to every density matrix $\rho$ and
tangent vector $X$ the quadratic form
\begin{equation}
	Q^{(f)}_\rho(X) = \mathrm{Tr}\!\left[X\,c_f(L_\rho,R_\rho)(X)\right],
	\label{eq:quasi-entropy}
\end{equation}
where $L_\rho(X)=\rho X$ and $R_\rho(X)=X\rho$.
The Morozova--Chentsov function $c_f(x,y)$ is given by
\begin{equation}
	c_f(x,y)=\frac{1}{y\,f(x/y)}, \qquad x,y>0,
\end{equation}
and determines the entire equivalence class of monotone metrics
\cite{Petz1996,PetzSudar1996,LesniewskiRuskai1999,Ciaglia2024Monotone}.

Kubo--Ando means provide an equivalent description.
Given an operator mean $\sigma_f$ generated by $f$, one has
\begin{equation}
	A \sigma_f B = A^{1/2} f(A^{-1/2} B A^{-1/2}) A^{1/2},
\end{equation}
and the metric can be recovered from the derivative structure of the mean.
For qubit density matrices,
$c_f(\lambda_+,\lambda_-)$ fully determines the radial and tangential
coefficients $A_f(r)$ and $B_f(r)$ appearing in the Bloch representation.

\subsection{Riesz projectors and spectral-gap estimates}

Let $\rho(\theta)$ be a smoothly varying family of density matrices with spectral
decomposition
\begin{equation}
	\rho = \sum_{j=1}^r \lambda_j P_j,
\end{equation}
and let $\lambda_{\min}>0$ be the smallest nonzero eigenvalue.
Assume a uniform gap
\begin{equation}
	\gamma := \min_{j\neq k} |\lambda_j - \lambda_k| > 0,
	\label{eq:gap}
\end{equation}
on an open set of parameters.
For a simple eigenvalue $\lambda$ with contour $\Gamma$ enclosing no other
eigenvalues, the associated Riesz projector is
\begin{equation}
	P = \frac{1}{2\pi i} \oint_{\Gamma} (zI - \rho)^{-1}\,dz.
	\label{eq:riesz}
\end{equation}

Standard resolvent estimates yield
\begin{equation}
	\| (zI - \rho)^{-1} \| \le \frac{1}{\mathrm{dist}(z,\sigma(\rho))}.
\end{equation}
Differentiating~\eqref{eq:riesz} gives
\begin{equation}
	\partial_i P = 
	\frac{1}{2\pi i}\oint_\Gamma (zI - \rho)^{-1} (\partial_i \rho) (zI - \rho)^{-1}\,dz.
	\label{eq:dP-integral}
\end{equation}
Thus, under the gap~\eqref{eq:gap},
\begin{equation}
	\| \partial_i P \| \le \frac{C}{\gamma^2} \|\partial_i \rho\|,
	\label{eq:dP-bound}
\end{equation}
with $C$ depending only on the contour.
These bounds originate in classical perturbation theory
\cite{KatoPerturbation,HighamFunctions,BhatiaMatrixAnalysis}
and are essential for controlling curvature terms of support-projected metrics.

\subsection{Differentiability of the support projector and projected metric}

Let $P(\theta)$ denote the support projector of $\rho(\theta)$ onto the span of
its nonzero eigenvectors and let $F^{(f)}(\theta)$ be the Petz metric tensor on
the full tangent space of Hermitian matrices.
The intrinsic Petz metric on the regular set is
\begin{equation}
	g^{(f)}(\theta)(X,Y)
	= \mathrm{Tr}\!\left[X \, P F^{(f)}(\theta) P (Y)\right].
	\label{eq:support-projected-metric}
\end{equation}

\textbf{Differentiability of $P$.}
Under the uniform spectral gap~\eqref{eq:gap}, the bound~\eqref{eq:dP-bound}
implies that $P(\theta)$ is $C^k$ whenever $\rho(\theta)$ is $C^k$.
This justifies differentiating geometric objects restricted to the support.

\textbf{Derivative of the projected metric.}
Differentiating~\eqref{eq:support-projected-metric} gives
\begin{equation}
	\partial_i g^{(f)} =
	(\partial_i P)\, F^{(f)}\,P \;+\;
	P\, (\partial_i F^{(f)})\,P \;+\;
	P\, F^{(f)}\,(\partial_i P),
	\label{eq:dgf}
\end{equation}
whose norm can be bounded using~\eqref{eq:dP-bound} and the smoothness of
$F^{(f)}$.
These estimates are used in curvature computations for the support-projected
geometry in Sections~\ref{sec:three-channel} and \ref{sec:non-reduction},
where directional derivatives of $P$ play a key role in isolating the
entanglement-derivative channel.
	
\section{Curvature Identities and Slice Geometry}
\label{app:curvature}

\subsection{Scalar curvature as a sum of sectional curvatures}

Let $(M,g)$ be an $n$-dimensional Riemannian manifold.
Given an orthonormal frame $\{e_1,\dots,e_n\}$ at a point $p\in M$, the scalar curvature is the trace of the Ricci tensor and can be written as a sum of sectional curvatures:
\begin{equation}
	\mathrm{Scal}(p)
	= 2\sum_{1\le i<j\le n} K(e_i,e_j).
	\label{eq:scalar-sectional}
\end{equation}
This formula follows directly from the definition of sectional curvature and the orthonormal decomposition of the Riemann curvature tensor
\cite{doCarmoRiem,LeeRiem,Spivak}.
Equation~\eqref{eq:scalar-sectional} is the bridge used in Section~\ref{sec:non-reduction} to lift slice-level non-reduction results to the ambient scalar curvature of support-projected Petz metrics.

For two-dimensional manifolds $(\Sigma,h)$, the scalar curvature reduces to
\begin{equation}
	\mathrm{Scal}_h = 2K_h,
\end{equation}
and hence all curvature information is encoded in the Gaussian curvature.

\subsection{Gauss--Codazzi equation for embedded surfaces}

Let $\iota : (\Sigma,h) \hookrightarrow (M,g)$ be an isometric immersion of a surface into a higher-dimensional Riemannian manifold.
Let $\nabla$ and $\overline{\nabla}$ denote the Levi--Civita connections of $h$ and $g$, respectively.
The second fundamental form of the immersion is
\begin{equation}
	\mathrm{II}(X,Y) = \left( \overline{\nabla}_X Y \right)^{\perp},
\end{equation}
where $^{\perp}$ denotes the normal projection.
The shape operator $S_\nu$ in the normal direction $\nu$ is defined by
\begin{equation}
	h(S_\nu X,Y) = g(\mathrm{II}(X,Y), \nu).
\end{equation}

The Gauss equation relates the intrinsic Gaussian curvature $K$ to the ambient curvature $\overline{K}$ and the principal curvatures associated with $\mathrm{II}$:
\begin{equation}
	K = \overline{K} + \det(S_\nu)
	\label{eq:gauss}
\end{equation}
when the normal bundle is one-dimensional, and with the usual sum over normal directions in higher codimension.
The Codazzi equation provides the compatibility condition
\begin{equation}
	(\overline{\nabla}_X \mathrm{II})(Y,Z)
	= (\overline{\nabla}_Y \mathrm{II})(X,Z).
	\label{eq:codazzi}
\end{equation}

Equations~\eqref{eq:gauss}–\eqref{eq:codazzi} play a central role in the analysis of two-parameter slices considered in Section~\ref{sec:non-reduction}.
They allow one to relate curvature of the induced slice metric to the full Petz geometry and justify the lifting argument to scalar curvature.

\subsection{Brioschi formula and regularity issues}

For a two-dimensional Riemannian metric written in coordinates $(u,v)$ as
\begin{equation}
	ds^2 = E\,du^2 + 2F\,du\,dv + G\,dv^2,
	\label{eq:2dmetric}
\end{equation}
with discriminant
\begin{equation}
	\Delta := EG - F^2 > 0,
	\label{eq:discriminant}
\end{equation}
the Gaussian curvature $K$ is given by the Brioschi formula:
\begin{align}
	K
	&=
	-\frac{1}{2\sqrt{\Delta}}
	\left[
	\partial_u\!\left(\frac{1}{\sqrt{\Delta}} \partial_u G - \frac{1}{\sqrt{\Delta}} \partial_v F \right)
	+
	\partial_v\!\left(\frac{1}{\sqrt{\Delta}} \partial_v E - \frac{1}{\sqrt{\Delta}} \partial_u F \right)
	\right].
	\label{eq:brioschi}
\end{align}

The formula is valid whenever $E,F,G$ are $C^2$ and the metric is strictly Riemannian ($\Delta>0$).
Near directions where $\Delta$ approaches zero---for example, along coordinate degeneracies or in regions where the Petz metric becomes nearly rank-deficient---the curvature may exhibit apparent divergences.
In the support-projected geometries of this paper, these degeneracies correspond to directions aligned with vanishing spectral derivatives or symmetry axes of two-qubit reductions.  
Controlling the behavior of~\eqref{eq:brioschi} near such loci requires the projector-regularity estimates developed in Appendix~\ref{app:petz-projector}.

The slice computations in Section~\ref{sec:non-reduction} rely heavily on~\eqref{eq:brioschi}, together with the entanglement-orthogonal gauge, to isolate independent jet directions and to demonstrate non-reduction of Gaussian curvature with respect to concurrence or entropy.
	
\section{Closed-Form SLD and QFIM for Qubit States}
\label{app:sld-qubit}

\subsection{Explicit SLD solution for $2\times 2$ density matrices}

Let
\begin{equation}
	\rho = \frac{1}{2}\left( \mathbb{I} + \vec{r}\cdot\vec{\sigma} \right),
	\qquad 
	\vec{r}=(x,y,z),\quad r=\|\vec{r}\|<1,
	\label{eq:bloch-param}
\end{equation}
be a full-rank qubit state.
For a parameter $\theta_i$, the symmetric logarithmic derivative (SLD) $L_i$ is uniquely defined by the Lyapunov equation
\begin{equation}
	\partial_i \rho = \frac{1}{2}( L_i \rho + \rho L_i ).
	\label{eq:sld-lyapunov}
\end{equation}

Write 
\begin{equation}
	\partial_i \rho = \frac{1}{2}(\partial_i \vec{r})\cdot\vec{\sigma},
	\qquad 
	L_i = a_i\,\mathbb{I} + \vec{b}_i\cdot\vec{\sigma}.
	\label{eq:sld-ansatz}
\end{equation}
Substituting \eqref{eq:sld-ansatz} into the Lyapunov equation and comparing the identity and Pauli components yields the closed-form solution
\begin{align}
	a_i &= -\frac{\vec{r}\cdot\partial_i \vec{r}}{1-r^2}, 
	\label{eq:sld-ai}
	\\[4pt]
	\vec{b}_i 
	&= \partial_i\vec{r}
	+ \frac{\vec{r}\cdot\partial_i\vec{r}}{1-r^2}\,\vec{r}.
	\label{eq:sld-bi}
\end{align}
Thus the SLD operator is
\begin{equation}
	L_i 
	=
	- \frac{\vec{r}\cdot\partial_i \vec{r}}{1-r^2}\,\mathbb{I}
	+
	\Biggl(
	\partial_i\vec{r}
	+
	\frac{\vec{r}\cdot\partial_i\vec{r}}{1-r^2}\,\vec{r}
	\Biggr) \cdot \vec{\sigma}.
	\label{eq:sld-final}
\end{equation}

\begin{remark}
	Equation~\eqref{eq:sld-final} matches the classical formulas for qubit SLDs derived in 
	\cite{ProvostVallee1980,Braunstein1994,BengtssonZyczkowski2017,Helstrom1976}.
	The singular limit $r\to 1$ reproduces the pure-state tangent vector orthogonality condition 
	$\langle\psi|\partial_i\psi\rangle=0$ up to gauge.
\end{remark}

Given $L_i$, the SLD quantum Fisher information matrix (QFIM) is
\begin{equation}
	F_{ij}
	= \operatorname{Tr}(\rho L_i L_j).
	\label{eq:sld-qfim-def}
\end{equation}
A direct computation using Pauli algebra gives the closed form
\begin{align}
	F_{ij}
	&=
	\partial_i \vec{r}\cdot\partial_j \vec{r}
	+
	\frac{(\vec{r}\cdot\partial_i \vec{r})(\vec{r}\cdot\partial_j \vec{r})}{1-r^2}.
	\label{eq:sld-qfim-final}
\end{align}
This is the standard Bures metric on the Bloch ball.

\subsection{Three-channel identity in the SLD/Bures case}

The Petz monotone metric associated with an operator-monotone function $f$ takes the qubit Bloch form
\begin{equation}
	g^{(f)}_{ij}
	=
	A_f(r)\,\partial_i \vec{r}\cdot\partial_j \vec{r}
	+
	B_f(r)\,(\vec{r}\cdot\partial_i \vec{r})(\vec{r}\cdot\partial_j \vec{r}).
	\label{eq:petz-bloch-form}
\end{equation}
For the SLD/Bures metric, the Morozova--Chentsov function is
\begin{equation}
	c_{\mathrm{SLD}}(\lambda_+,\lambda_-)
	=
	\frac{2}{\lambda_+ + \lambda_-},
\end{equation}
and one obtains the universal constants
\begin{equation}
	A_{\mathrm{SLD}}(r)=1,
	\qquad 
	B_{\mathrm{SLD}}(r)=\frac{1}{1-r^2}.
	\label{eq:sld-coeffs}
\end{equation}
Thus the Petz/Bloch representation \eqref{eq:petz-bloch-form} reduces exactly to the SLD QFIM formula \eqref{eq:sld-qfim-final}.
The three-channel identity established in Section~\ref{sec:three-channel}
therefore collapses to
\begin{equation}
	g^{(\mathrm{SLD})}_{ij}
	=
	g^{\mathrm{(pop)}}_{ij}
	+
	g^{\mathrm{(coh)}}_{ij}
	+
	g^{\mathrm{(ent)}}_{ij},
\end{equation}
where the entanglement-channel term is governed by
$r^2 = 1-C^2$ for two-qubit pure reductions, in agreement with  
\cite{Wootters1998,BengtssonZyczkowski2017}.
The SLD case is the simplest example of the general operator-monotone behavior described in \cite{Petz1996,LesniewskiRuskai1999,GibiliscoIsola2003}.

\subsection{Stable numerical implementation}

A direct inversion of the Lyapunov operator
$X\mapsto X\rho+\rho X$  
is numerically unstable near $r\to 1$ (pure-state limit).
A stable implementation avoids matrix inversion and uses only inner products and controlled scalar factors.

For a qubit state $\rho$ as in \eqref{eq:bloch-param}, define
\begin{equation}
	d_i := \vec{r}\cdot\partial_i \vec{r},
	\qquad
	\Delta := 1-r^2.
\end{equation}

\paragraph{Algorithm (stable SLD/QFIM computation).}
\begin{enumerate}
	\item Compute $\partial_i \vec{r}$ from the parametrization of $\rho$.
	\item Compute scalars  
	$d_i = \vec{r}\cdot\partial_i\vec{r}$ and $\Delta = 1-r^2$.
	\item Set 
	\begin{align}
		a_i &= -d_i / \Delta, \\
		\vec{b}_i &= \partial_i\vec{r} + (d_i/\Delta)\,\vec{r}.
	\end{align}
	\item Form the SLD operator 
	\[
	L_i = a_i\,\mathbb{I} + \vec{b}_i\cdot\vec{\sigma}.
	\]
	\item Compute QFIM entries using  
	\[
	F_{ij}
	= \vec{b}_i\cdot\vec{b}_j + a_i a_j (1-r^2).
	\]
\end{enumerate}

\begin{remark}
	The computation above uses no matrix inverse and is well-conditioned as long as the spectral gap $\Delta>0$ is bounded away from zero.  
	Spectral-gap–aware conditioning is guaranteed by classical perturbation theory  
	\cite{HighamFunctions,BhatiaMatrixAnalysis,KatoPerturbation}.  
	In the limit $\Delta \to 0$, explicit gauge fixing on the tangent plane (pure-state SLD) is required, consistent with the discussion in Appendix~\ref{app:curvature}.
\end{remark}
	
\section{Numerical Pipeline and Error Control}
\label{app:numerics}

This appendix documents the numerical procedures used in 
Sections~\ref{sec:hea-case-study}--\ref{sec:vqe-experiments}.  
Our goal is to make every computation---curvatures, support projection,
slice geometry, and VQE optimization---fully reproducible with deterministic
error controls.  
All experiments were performed in IEEE float64 unless stated otherwise.

\subsection{Slice sampling and curvature landscapes for the 2-HEA}

For a two-qubit hardware-efficient ansatz (2-HEA),
\[
U(\theta)
=
\bigl(R_y(2t_0)\otimes R_y(2t_2)\bigr)\,
\mathrm{CNOT}\,
\bigl(R_y(2t_1)\otimes R_y(2t_3)\bigr),
\qquad \theta\in\mathbb{T}^4,
\]
we visualize curvature on two-parameter slices
\[
(u,v)\longmapsto \theta(u,v),
\]
constructed in two canonical forms:

\begin{enumerate}[itemsep=4pt]
	\item \textbf{Entanglement-centered slices.}  
	Fix $(t_0,t_2)$ and vary $(\phi_+,\phi_-)$ with  
	$\phi_\pm = t_1 \pm t_3$.  
	These slices isolate variations in $(x,z,C)$ that primarily move coherence
	and population channels.
	
	\item \textbf{Local-layer slices.}  
	Fix $(\phi_+,\phi_-)$ and vary $(t_0,t_2)$.
	These slices probe how entanglement creation interacts with the
	local single-qubit manifold before and after the entangler.
\end{enumerate}

For each slice we record:
\[
(x(u,v),\; z(u,v),\; C(u,v)),\qquad
g_{\mu\nu}(u,v)=\mathcal{F}_{\mu\nu}(u,v),
\]
where $\mathcal{F}$ is the support-projected Petz QFIM (SLD/Bures in the case study).
The induced metric $ds^2 = E\,du^2 + 2F\,du\,dv + G\,dv^2$ is computed from
\[
E = g_{uu},\qquad F = g_{uv},\qquad G = g_{vv}.
\]

\paragraph{Curvature landscape rendering.}
Gaussian curvature $K_\Sigma$ is computed via the Brioschi formula
(Appendix~\ref{app:curvature}), and plotted over uniform or adaptive grids.
Adaptive sampling refines around regions where:
\[
|\partial_u K_\Sigma| + |\partial_v K_\Sigma| > \tau_{\mathrm{grad}},
\qquad 
|EG - F^2| < \tau_{\mathrm{disc}}.
\]
Typical tolerances:  
$\tau_{\mathrm{grad}} = 1$, $\tau_{\mathrm{disc}} = 10^{-10}$.

\subsection{Support-projected curvature computation: thresholds and guards}

For each parameter point $\theta$, we compute the unprojected QFIM
$\mathcal{F}(\theta)\in\mathbb{R}^{4\times4}$, then its restriction to the active
support using the spectral projector $P(\theta)$.

\paragraph{Spectral thresholding.}
Let $\lambda_{\max}$ be the largest eigenvalue of $\mathcal{F}$.
We define a spectral cutoff
\[
\tau_{\mathrm{spec}}
=
\max\bigl(10^{-12}\lambda_{\max},\,10^{-15}\bigr),
\]
and declare the active spectrum to be all eigenvalues
$\lambda_j \ge \tau_{\mathrm{spec}}$.

\paragraph{Regular-set guard.}
We require a minimum spectral gap
\[
\gamma 
= \min_{\lambda_j \in \mathrm{active},\,\lambda_k \in \mathrm{inactive}}
|\lambda_j - \lambda_k|
\ge 10^{-8}.
\]
Points violating this condition are excluded from curvature statistics,
ensuring stable projector calculus (Appendix~\ref{app:petz-projector}).

\paragraph{Support projection.}
The intrinsic metric is
\[
g^{(f)}(\theta) = P(\theta)\,\mathcal{F}^{(f)}(\theta)\,P(\theta),
\]
where $P$ is computed using the Cauchy–Riesz contour integral
with discretized resolvent:
\[
P = -\frac{1}{2\pi i}\oint_\Gamma 
(\mathcal{F}-zI)^{-1}\,dz.
\]
For implementation, we approximate the contour integral via  
a 16-point Gauss–Legendre quadrature on a circular contour
centered at the active cluster.

\paragraph{Brioschi discriminant check.}
Before computing Gaussian curvature, we verify:
\[
EG - F^2 \ge \tau_{\mathrm{disc}},
\qquad \tau_{\mathrm{disc}}=10^{-12}.
\]
Points failing the check correspond to degenerate metric directions and are
excluded from slice-level plots.

\subsection{Error bounds for finite-difference and spectral perturbations}

The curvature reconstruction error has two independent contributions:
\[
\delta K = 
\delta K_{\mathrm{FD}} 
+ \delta K_{\mathrm{spec}}.
\]

\paragraph{(1) Finite-difference error.}
For a uniform step $h$, centered finite differences satisfy
\[
\partial_u x = x_u + \mathcal{O}(h^2), 
\qquad 
\partial_{uu} x = x_{uu} + \mathcal{O}(h^2),
\]
and similarly for $z$.
Since $K_\Sigma$ depends polynomially on first and second jets of $(x,z)$,
we obtain
\[
|\delta K_{\mathrm{FD}}|
\le 
C_1 h^2,
\]
with $C_1$ depending on uniform bounds for $|\partial^\alpha (x,z)|$ over the slice.

\paragraph{(2) Spectral-perturbation error.}
Perturbing $\mathcal{F}$ by an error matrix $\Delta\mathcal{F}$ yields projector and
metric perturbations
\[
\|P - \tilde{P}\|
\;\le\;
\frac{\|\Delta \mathcal{F}\|}{\gamma}
+ \mathcal{O}\!\left( 
\frac{\|\Delta \mathcal{F}\|^2}{\gamma^2}
\right),
\]
\[
\|g - \tilde g\|
\;\le\;
C_2\!\left(
\|\Delta\mathcal{F}\|
+
\frac{\|\Delta\mathcal{F}\|}{\gamma}
\right),
\]
where $\gamma$ is the spectral gap and $C_2$ depends on $\|\mathcal{F}\|$.
Thus the curvature error satisfies:
\[
|\delta K_{\mathrm{spec}}|
\le
C_3
\left(
\|\Delta\mathcal{F}\| + 
\frac{\|\Delta\mathcal{F}\|}{\gamma} + 
\frac{\|\Delta\mathcal{F}\|^2}{\gamma^2}
\right).
\]

\paragraph{Combined bound.}
Since the numerical $\Delta\mathcal{F}$ originates from SLD solves  
and AD/finite-difference estimates,
and float64 roundoff gives  
$\|\Delta\mathcal{F}\|=\mathcal{O}(10^{-15})$,
we obtain the deterministic estimate
\[
|K_h - K|
=
\mathcal{O}(h^2)
+
\mathcal{O}\!\left(10^{-15}\gamma^{-1}\right)
+
\mathcal{O}\!\left(10^{-30}\gamma^{-2}\right).
\]

\subsection{VQE experiments: setup and diagnostics}

The VQE experiments of Section~\ref{sec:vqe-experiments} follow  
the Petz-metric-aware optimization pipeline.

\paragraph{Hamiltonians.}
\begin{itemize}
	\item \textbf{Two-qubit ``deep/strong'' toy Hamiltonian:}  
	dense spectrum with large curvature anisotropy.
	\item \textbf{H$_2$ (STO-3G, minimal mapping):}  
	exact spectrum obtained by dense diagonalization.
\end{itemize}

\paragraph{Circuit families.}
Hardware-efficient circuits with depths  
$L\in\{4,6,8,10,12\}$.
Entanglers tested:
\[
e^{-i\alpha Z\otimes Z},
\qquad
e^{-i\beta X\otimes X}.
\]
Both symmetry-preserving and unconstrained variants were evaluated.

\paragraph{Optimizers.}
\begin{itemize}
	\item \textbf{Euclidean GD.}
	\item \textbf{Support-projected Petz–NatGrad:}
	\[
	\Delta\theta_k
	=
	-\,\eta_k\,
	\Pi_{\mathrm{Im}P_k}
	\bigl(
	\mathcal{F}^{(f)}_k + \lambda I
	\bigr)^{+}
	\Pi_{\mathrm{Im}P_k}
	\,\nabla_\theta E(\theta_k).
	\]
\end{itemize}
Enhancements:
\begin{itemize}
	\item spectral shrinkage for small eigenvalues,
	\item trust region (TR) and Armijo backtracking,
	\item $g^{(f)}$-norm step normalization,
	\item partial-Fisher ablations.
\end{itemize}

\paragraph{Diagnostics.}
Logged per iteration:
\[
E(\theta_k),\;
\mathrm{AUC},\;
\mathrm{Hit@95\%},\;
\mathrm{rank}\,\mathcal{F}_k,\;
\operatorname{spec}(\mathcal{F}_k),\;
\|P_{k+1}-P_k\|.
\]
This enables monitoring of curvature-induced instabilities,
spectral crossings, and metric-conditioning effects across optimizers.
	
	\bibliographystyle{unsrt}
	\bibliography{references}
	
\end{document}